\documentclass[aps,prc,nofootinbib]{revtex4-2}
\usepackage{amsmath,amssymb}
\usepackage{dsfont}
\usepackage{bm} 
\usepackage{graphicx}

\newcommand{\be}{\begin{equation}}
\newcommand{\ee}{\end{equation}}
\newcommand{\ba}{\begin{eqnarray}}
\newcommand{\ea}{\end{eqnarray}}
\newcommand{\ban}{\begin{eqnarray*}}
\newcommand{\ean}{\end{eqnarray*}}
\newcommand \nn {\nonumber}

\begin{document}

\title{Stability of Initial Glasma Fields}

\author{Sylwia Bazak$^1$\footnote{e-mail: sylwia.bazak@gmail.com} and Stanis\l aw Mr\' owczy\' nski$^{1,2}$\footnote{e-mail: stanislaw.mrowczynski@ncbj.gov.pl}}

\affiliation{$^1$Institute of Physics, Jan Kochanowski University, ul. Uniwersytecka 7, PL-25-406 Kielce, Poland 
\\
$^2$National Centre for Nuclear Research, ul. Pasteura 7,  PL-02-093 Warsaw, Poland}

\date{December 17, 2023}

\begin{abstract}

A system of gluon fields produced in the earliest phase of relativistic heavy-ion collisions, which is called {\it glasma}, can be described in terms of classical fields. Initially there are chromoelectric and chromomagnetic fields along the collision axis. A linear stability analysis of these fields is performed, assuming that the fields are space-time uniform and using the SU(2) gauge group. We apply Milne coordinates and the gauge condition which are usually used in studies of glasma. The chromoelectric field is in the Abelian configuration with the corresponding potential linearly depending on coordinates but the chromomagnetic field is in the nonAbelian configuration generated by the potential of non-commuting components. The chromomagnetic field is found to be unstable and the growth rate of the unstable mode is derived. Our findings are critically debated and confronted with the numerical simulations by Romatschke and Venugopalan who found that the evolving glasma is unstable due to the Weibel instability well-known in electromagnetic plasma. 

\end{abstract}

\maketitle

\section{Introduction}

The Color Glass Condensate (CGC) approach, see {\it e.g.} the review articles \cite{Iancu:2003xm,Gelis:2012ri}, is commonly applied to study the early phase of relativistic heavy-ion collisions at the highest accessible energies at RHIC and the LHC. Within the approach, valence quarks of the colliding nuclei act as color charge sources of long wavelength chromodynamic fields. Such a system of gluon fields created in the nuclear collision is called {\it glasma} and the fields can be approximately treated as classical because of their large occupation numbers. Initially the system is dominated by the chormoelectric and chromomagnetic fields parallel to the beam axis. Later on transverse fields show up and the system evolves towards thermodynamic equilibrium.  

Numerical simulations of the evolving glasma \cite{Romatschke:2005pm,Romatschke:2006nk} showed that the system is unstable. The exponentially growing mode was identified with the Weibel instability \cite{Wei59}, also called the {\it filamentation}, which is well-known in physics of electromagnetic plasma. A relevance of the chromodynamic Weibel instability for quark-gluon plasma produced in relativistic heavy-ion collisions was argued long ago \cite{Mrowczynski:1993qm} and studied in detail further on, see the review \cite{Mrowczynski:2016etf}. A sufficient condition for an occurrence of the instability is an anisotropy of momentum distribution of plasma charged constituents. They generate magnetic field which in turn affects their motion and, as explained in detail in \cite{Mrowczynski:2016etf}, an energy is transferred from the particles to the exponentially growing field. The interplay of particles and fields is crucial for the Weibel instability, but, in principle, there are no particles in the glasma -- quasiparticles are expected to appear later on as the system evolves towards thermodynamic equilibrium. Nevertheless, high-frequency modes of classical fields are often treated as quasiparticles and such a treatment was behind the interpretation of the instability found in \cite{Romatschke:2005pm,Romatschke:2006nk} as the Weibel mode. 

In the early days of QCD various configurations of classical chromodynamic fields were found unstable, see \cite{Mandula:1976uh,Mandula:1976xf,Chang:1979tg,Sikivie:1979bq,Tudron:1980gq}. This observation was a starting point of a whole series of papers \cite{Fukushima:2007yk,Iwazaki:2007es,Iwazaki:2008xi,Fujii:2008dd,Fujii:2009kb} where the problem of glasma stability was studied. Uniform chromoelectric and chromomagnetic fields along the beam direction were considered and it was argued \cite{Iwazaki:2007es,Iwazaki:2008xi,Fujii:2008dd,Fujii:2009kb} that the instability found in the simulations \cite{Romatschke:2005pm,Romatschke:2006nk} is not the Weibel but rather Nielsen-Olesen instability \cite{Nielsen:1978rm} of spin 1 charged particles circulating in a uniform chromomagnetic field. There was also considered \cite{Fujii:2008dd} a possible role of the vacuum instability caused by a strong chromoelectric field which generates particle-antiparticle pairs due to the Schwinger mechanism \cite{Schwinger:1951nm}. 

An important aspect of the glasma chromomagnetic field was missed in the studies \cite{Iwazaki:2007es,Iwazaki:2008xi,Fujii:2008dd,Fujii:2009kb}. Uniform chromoelectric and chromomagnetic fields are generated by the potentials which are either of single color and linearly depend on coordinates or by uniform multi-color potentials the components of which do not commute with each other. We call the former configurations -- fully analogous to those known from electrodynamics -- as Abelian and the latter ones as nonAbelian. It is important to note that the Abelian and nonAbelian configurations are physically nonequivalent as they cannot be related to each other by a gauge transformation \cite{Brown:1979bv}. We also note that although the uniform chromomagnetic field is unstable in both Abelian and nonAbelian configurations, characteristics of the unstable modes are different \cite{Bazak:2021xay}. We will show in Sec.~\ref{sec-glasma-fields} that the initial chromoelectric glasma field is in the Abelian configuration but the chromomagnetic one is the nonAbelian one. In the studies \cite{Iwazaki:2007es,Iwazaki:2008xi,Fujii:2008dd,Fujii:2009kb} the chromoelectric and chromomagnetic fields were both assumed to be in the Abelian configuration.

We have undertaken an effort to systematically study a stability of classical glasma fields. We have performed \cite{Bazak:2021xay,Bazak:2022khg} a linear stability analysis of space-time uniform chromoelectric and chromomagnetic fields in the Abelian configurations, which were studied repeatedly starting with the works \cite{Chang:1979tg,Sikivie:1979bq}, and in the nonAbelian configurations which were briefly discussed in \cite{Tudron:1980gq}, see also \cite{Berges:2011sb, Vachaspati:2022ktr,Pereira:2022lbl}, but not systematically studied. We have derived a complete spectrum of small fluctuations around the background fields which obey the linearized Yang-Mills equations. The spectra of Abelian and nonAbelian configurations are similar but different and they both include unstable modes. We have also discussed parallel chromoelectric and chromomagnetic fields, which occur simultaneously. 

The aim of this work is to perform a stability analysis of uniform chromodynamic fields which are relevant for the initial glasma state and confront the results with the simulations \cite{Romatschke:2005pm,Romatschke:2006nk}. We use the Milne -- also known as co-moving -- coordinates and we apply a specific Fock-Schwinger gauge condition which is usually used in studies of glasma. We note that in our previous works \cite{Bazak:2021xay,Bazak:2022khg} we have been using Minkowski coordinates and the background gauge which is convenient to compare various systems under consideration.  

Our paper is organized as follows. In Sec.~\ref{sec-glasma-fields} we discuss an Ansatz \cite{Kovner:1995ts} which determines a  structure of glasma field. We show that the initial chromoelectric field is in the Abelian configuration while the chromomegntic field is generated by the genuine nonAbelian potential. Yang-Mills equations in Milne coordinates are discussed in Sec.~\ref{sec-YM-Milne} where we also derive the equations linearized in a small deviation from the background potential. In Secs.~\ref{sec-stab-B} and \ref{sec-stab-E} a linear stability analysis is performed of the initial chromomagnetic and chromoelectric fields, respectively. The initial glasma fields are treated as stationary but to check a reliability of the assumption we consider  in Sec.~\ref{sec-glasma-evolution} the evolution of glasma fields, using the proper time expansion proposed in \cite{Fries:2006pv} and developed in \cite{Chen:2015wia}. In the closing section~\ref{sec-discussion} our findings are critically debated and discussed in the context of the simulations \cite{Romatschke:2005pm,Romatschke:2006nk,Ruggieri:2017ioa}. 

Throughout the paper we use Minkowski, light-cone and Milne coordinates
\be
(t,{\bf x}_\perp,z), ~~~~~~~~~(x^+,{\bf x}_\perp, x^-),~~~~~~~~~(\tau,{\bf x}_\perp,\eta),
\ee
where $x^\pm \equiv (t\pm z)/\sqrt{2}$, $\tau \equiv\sqrt{t^2-z^2}=\sqrt{2x^+ x^-}$, $\eta \equiv\ln(x^+/x^-)/2$ and ${\bf x}_\perp = (x,y)$. The indices $\mu, \nu$, which label coordinates of spacetime, are $\mu, \nu = t, x, y, z$ in case of Minkowski coordinates and $\mu, \nu = \tau, x, y, \eta$ in case of Milne ones. The indices $\alpha, \beta = x, y, z$ and $i,j = x, y$ label, respectively, the Cartesian spatial coordinates and those of $x\!-\!y$ plane which is transverse to the beam direction along the axis $z$. The space-time metric tensor $g_{\mu \nu}$ is diagonal and the diagonal elements are $(+1,-1,-1,-1)$ in Minkowski and $(+1,-1,-1,-\tau^2)$ in Milne coordinates. The indices $a, b = 1, 2, \dots N_c^2 -1$ numerate color components in the adjoint representation of SU($N_c$) gauge group. We neglect henceforth the prefix `chromo' when referring to chromoelectric or chromomagnetic fields. Since we study chromodynamics only, this should not be confusing.

\section{Glasma Fields}
\label{sec-glasma-fields}

We consider a collision of two heavy ions moving  towards each other along the $z$-axis with the speed of light and colliding at $t=z=0$. The vector potential of the gluon field is described with the Ansatz \cite{Kovner:1995ts} 
\ba
\nn
A^+(x) &=& \Theta(x^+)\Theta(x^-) \, x^+ \alpha(\tau,{\bf  x}_\perp) ,
\\[2mm] \label{ansatz}
A^-(x) &=& -\Theta(x^+)\Theta(x^-) \, x^- \alpha(\tau,{\bf  x}_\perp) ,
\\[2mm] \nn
A^i(x) &=& \Theta(x^+)\Theta(x^-) \alpha_\perp^i(\tau,{\bf  x}_\perp)
+\Theta(-x^+)\Theta(x^-) \beta_1^i({\bf  x}_\perp)
+\Theta(x^+)\Theta(-x^-) \beta_2^i({\bf  x}_\perp) ,
\ea
where the functions $\beta_1^i(x^-,{\bf  x}_\perp)$ and $\beta_2^i(x^+,{\bf  x}_\perp)$ represent the pre-collision potentials and the functions $\alpha(\tau,{\bf  x}_\perp)$ and $\alpha_\perp^i(\tau,{\bf  x}_\perp)$ give the post-collision potentials, $i,j=x,y$.

Components of the potential (\ref{ansatz}) in the forward light-cone ($x^\pm \ge 0$), where glasma is present, are
\ba
\left\{
\begin{array}{l}
A^t(x) = z \,\alpha(\tau,{\bf  x}_\perp) ,
\\[2mm] 
A^z(x) =  t \, \alpha(\tau,{\bf  x}_\perp) ,
\\[2mm] 
A^i(x) = \alpha_\perp^i(\tau,{\bf  x}_\perp) ,
\end{array} \right. 
~~~~~~~~~~~~~~~~~~~~~~~~
\left\{
\begin{array}{l}
A^\tau(x) =  0,
\\[2mm] \label{potential-components}
A^\eta(x) = \alpha(\tau,{\bf  x}_\perp) ,
\\[2mm] 
A^i(x) = \alpha_\perp^i(\tau,{\bf  x}_\perp) ,
\end{array} \right. 
\ea
in Minkowski and Milne coordinates, respectively. 

The potential (\ref{ansatz}) satisfies the specific Fock-Schwinger gauge condition which written in the Minkowski, light-cone and Milne coordinate systems,  is
\be
\label{gauge-condition}
t A^t - z A^z = 0 , ~~~~~~~
x^-A^+ + x^+A^- = 0 , ~~~~~~~
A^\tau = 0 .
\ee

In the forward light-cone the vector potential satisfies the sourceless Yang-Mills equations but the sources enter through the boundary conditions that connect the pre- and post-collision potentials. The conditions read \cite{Kovner:1995ts}
\ba
\label{cond1}
\alpha^{i}_\perp(0,{\bf x}_\perp) &=& \beta^i_1 ({\bf x}_\perp) + \beta^i_2({\bf x}_\perp) ,
\\ [2mm]
\label{cond2}
\alpha(0,{\bf x}_\perp) &=& -\frac{ig}{2}[\beta^i_1 ({\bf x}_\perp),\beta^i_2 ({\bf x}_\perp)] .
\ea

In Minkowski coordinates the electric and magnetic fields are given as
\be
\label{EB-field}
E^\alpha = F^{\alpha t} , ~~~~~~~~~~
B^\alpha = \frac{1}{2} \epsilon^{\alpha \beta \gamma} F^{\gamma \beta} ,
\ee
where $\alpha, \beta, \gamma = x,y,z$, the strength tensor is
\be
\label{F1}
F^{\mu\nu}=\partial^\mu A^\nu - \partial^\nu A^\mu - ig [A^\mu, A^\nu] .
\ee
and $\epsilon^{\alpha \beta \gamma}$ is the Levi-Civita fully antisymmetric tensor.

Since $t = \tau \cosh \eta$ and $z = \tau \sinh \eta$, the components $A^t$ and $A^z$ given by Eq.~(\ref{potential-components}) vanish at $\tau=0$, as $ \alpha(\tau,{\bf  x}_\perp)$ is assumed to be regular at  $\tau=0$. However, the derivatives of $A^t$ and $A^z$ with respect of $z$ and $t$, respectively, are finite at $\tau=0$. Therefore, the only non-zero components of the electric and magnetic fields at $\tau=0$ are
\ba
\label{Ez-initial}
E({\bf  x}_\perp) \equiv E^z(0,{\bf  x}_\perp) &=& -2 \alpha(0,{\bf  x}_\perp) ,
\\ \label{Bz-initial}
B({\bf  x}_\perp) \equiv B^z(0,{\bf  x}_\perp) 
&=& - \partial_y \alpha_\perp^x(0,{\bf  x}_\perp) 
+ \partial_x \alpha_\perp^y(0,{\bf  x}_\perp) 
-ig[\alpha_\perp^y(0,{\bf  x}_\perp),\alpha_\perp^x(0,{\bf  x}_\perp)].
\ea

When the pre-collision potentials $\beta^i_1$ and $\beta^i_2$ are uniform, that is independent of ${\bf  x}_\perp$, the glasma initial potential is 
\ba
\nn
A_a^t &=& z \, \alpha_a = \frac{g}{2} \, z \, f^{abc} \beta^i_{1b} \beta^i_{2c} ,
\\[2mm] \label{initial-uniform-potential}
A_a^z &=& t \, \alpha_a = \frac{g}{2} \, t \, f^{abc} \beta^i_{1b} \beta^i_{2c} ,
\\[2mm] \nn
A_a^i &=&   \alpha_{\perp\,a}^i = \beta^i_{1a} + \beta^i_{2a} ,
\ea
which is written in the adjoint representation of the ${\rm SU}(N_c)$ gauge group. The potential generates the initial electric and magnetic fields which are
\ba
\label{Ez-initial-adj}
E_a   &=& - g  f^{abc} \beta^i_{1b}\beta^i_{2c} ,
\\[2mm] \label{Bz-initial-adj}
B_a  &=& -g \epsilon^{zij}  f^{abc} \beta^i_{1b} \beta^j_{2c} .
\ea

Eqs.~(\ref{initial-uniform-potential}) clearly show that the initial uniform electric field (\ref{Ez-initial-adj}) is in Abelian configuration that is it is generated by the potential components $A^t$ and $A^z$ linearly depending on $t$ and $z$ while the magnetic field (\ref{Bz-initial-adj}) is in nonAbelian configuration that is it is generated by the uniform and noncommuting potential components $A^x$ and $A^y$. 

One shows that the Abelian and nonAbelian configurations of the same uniform electric or magnetic field are physically nonequivalent by observing that the potential of Abelian configuration satisfies the Yang-Mills equations with vanishing color current, while the potential of nonAbelian configuration can solve the Yang-Mills equations only with an appropriately chosen color current. Since a nonzero current cannot be nulled by gauge transformation the Abelian and nonAbelian configurations are nonequivalent. We will return to this issue in Sec.~\ref{sec-stab-B}.

\section{Linearized Yang-Mills equations in Milne coordinates}
\label{sec-YM-Milne}

In curvilinear coordinates the Yang-Mills equations are
\be
\label{Y-M-eq-general}
\nabla_\mu F^{\mu \nu} = j^\nu ,
\ee
where $\nabla_\mu$ is the covariant derivative which includes Christoffel symbols and the gauge potential,
\be
F^{\mu \nu} \equiv \nabla^\mu A^\nu - \nabla^\nu A^\mu ,
\ee
and $j^\mu$ is the color current. In the adjoint representation of ${\rm SU}(N_c)$ group the covariant derivative acts on the vector potential $A^\mu_a$ as
\be
\nabla_\mu^{ab} A^\nu_b \equiv \partial_\mu A^\nu_a + \Gamma^\nu_{\mu \rho} A^\rho_a 
+ g f^{abc} A_\mu^b A^\nu_c ,
\ee
where $\Gamma^\nu_{\mu \rho}$ is the Christoffel symbol. We note that upper and lower color indices are not distinguished from each other. Since $\Gamma^\mu_{\nu \rho} = \Gamma^\mu_{\rho \nu}$, the strength tensor $F_{\mu \nu}^a$ equals
\be
\label{F-mu-nu-lower}
F_{\mu \nu}^a = (\nabla_\mu A_\nu - \nabla_\nu A_\mu)^a 
= \partial_\mu A_\nu^a - \partial_\nu A_\mu^a  + g f^{abc} A_\mu^b A_\nu^c .
\ee

To derive an explicit form of the Yang-Mills equations in Milne coordinates we use the formula  
\be
\nabla_\mu F^{\mu \nu} = \frac{1}{\sqrt{-\bar{g}}} D_\mu\big(\sqrt{-\bar{g}} \, F^{\mu \nu} \big) ,
\ee
where $\bar{g}$ is the determinant of a metric tensor which in case of Milne coordinates equals $\bar{g} = - \tau^2$ and $D_\mu$ is the covariant derivative which includes the gauge potential but no Christoffel symbol. The Yang-Mills equations can be written as
\be
\label{YM-eq-2}
(\nabla_\mu F^{\mu \nu})_a =   \frac{1}{\tau} \, \partial_\mu 
(\tau \,g^{\mu \rho} g^{\nu \sigma} F_{\rho \sigma}^a) 
+ g g^{\mu \rho} g^{\nu \sigma} f^{abc} A_\mu^b F_{\rho \sigma}^c =  j^\nu_a ,
\ee
where the derivatives $\partial_\mu$ but not $\partial^\mu$ are used and the strength tensor is given by Eq.~(\ref{F-mu-nu-lower}) which does not include Christoffel symbols.

Since the Yang-Mills equations play a central role in our analysis we write them explicitly in case of the ${\rm SU}(2)$ gauge group, when $f^{abc} = \epsilon^{abc}$, and the gauge condition is $A^\tau = 0$. The equations read
\ba
\label{YM-eq-tau-tau=0}
&& ~~~~ - \, \partial_x\partial_\tau A^x_a 
- \partial_y  \partial_\tau A^y_a 
- \, \frac{1}{\tau^2} \, \partial_\eta \partial_\tau (\tau^2 A^\eta_a) 
+ g \epsilon^{abc} A^x_b \partial_\tau A^x_c 
+ g \epsilon^{abc} A^y_b \partial_\tau A^y_c 
+ g \epsilon^{abc} A^\eta_b \partial_\tau (\tau^2 A^\eta_c) 
= j^\tau_a ,
\\[2mm] \nn
&& \partial_\tau^2 A^x_a 
+ \frac{1}{\tau} \, \partial_\tau A^x_a 
- \, \partial_y (\partial_y A^x_a - \partial_x A^y_a - g \epsilon^{abc} A^y_b A^x_c)
- \frac{1}{\tau^2} \, \partial_\eta 
(\partial_\eta A^x_a - \tau^2 \partial_x A^\eta_a - g \tau^2 \epsilon^{abc} A^\eta_b A^x_c) 
\\[2mm] \nn
&& ~~~~~~~~~~~~~~~~~~~~~~~~
+ g \epsilon^{abc} A^y_b (\partial_y A^x_c - \partial_x A^y_c) 
+ g \epsilon^{abc} A^\eta_b (\partial_\eta A^x_c - \tau^2 \partial_x A^\eta_c) 
\\[2mm] 
\label{YM-eq-x-tau=0}
&&~~~~~~~~~~~~~~~~~~~~~~~~~~~~~~~~~~~~~~~~~~~~~~~~~~~~
- g^2( A^y_b A^y_a A^x_b - A^y_b A^y_b A^x_a 
+ \tau^2 A^\eta_b A^\eta_a A^x_b - \tau^2 A^\eta_b A^\eta_b A^x_a) 
= j^x_a ,
\\[2mm] \nn
&& \partial_\tau^2 A^y_a  + \frac{1}{\tau} \, \partial_\tau A^y_a 
- \, \partial_x (\partial_x A^y_a - \partial_y A^x_a - g \epsilon^{abc} A^x_b A^y_c) 
- \frac{1}{\tau^2} \, \partial_\eta (\partial_\eta A^y_a - \tau^2 \partial_y A^\eta_a - g \tau^2  \epsilon^{abc} A^\eta_b A^y_c)  
\\[2mm] \nn
&& ~~~~~~~~~~~~~~~~~~~~~~~~
+ g \epsilon^{abc} A^x_b (\partial_x A^y_c - \partial_y A^x_c) 
+ g \epsilon^{abc} A^\eta_b (\partial_\eta A^y_c - \tau^2 \partial_y A^\eta_c) 
\\[2mm] 
\label{YM-eq-y-tau=0}
&& ~~~~~~~~~~~~~~~~~~~~~~~~~~~~~~~~~~~~~~~~~~~~~~~~~~~~
- g^2 (A^x_b A^x_a A^y_b - A^x_b A^x_b A^y_a
+ \tau^2 A^\eta_b A^\eta_a A^y_b - \tau^2 A^\eta_b A^\eta_b A^y_a) 
= j^y_a ,
\\[2mm] \nn
&& ~~ \partial_\tau^2 (\tau^2 A^\eta_a)
- \frac{1}{\tau}\partial_\tau (\tau^2 A^\eta_a) 
- \, \partial_x 
(\tau^2 \partial_x A^\eta_a - \partial_\eta A^x_a - g \tau^2 \epsilon^{abc} A^x_b A^\eta_c )  
- \, \partial_y 
(\tau^2 \partial_y A^\eta_a - \partial_\eta A^y_a - g \tau^2 \epsilon^{abc} A^y_b A^\eta_c )  
\\[2mm] \nn
&& ~~~~~~~~~~~~~~~~~~~~~~~~~~
+ g \epsilon^{abc} A^x_b (\tau^2 \partial_x A^\eta_c - \partial_\eta A^x_c) 
+ g \epsilon^{abc} A^y_b (\tau^2 \partial_y A^\eta_c - \partial_\eta A^y_c) 
\\[2mm] 
\label{YM-eq-eta-tau=0}
&& ~~~~~~~~~~~~~~~~~~~~~~~~~~~~~~~~~~~~~~~~~~~~~~~~~~~~
+ g^2 \tau^2 (- A^x_b A^x_a A^\eta_b + A^x_b A^x_b A^\eta_a
- A^y_b A^y_a A^\eta_b + A^y_b A^y_b A^\eta_a)
= \tau^2 j^\eta_a.
\ea

Now, we assume that the background potential $\bar{A}_a^\mu$ solves the Yang-Mills equations and we consider small fluctuations $a_a^\mu$ around $\bar{A}_a^\mu$. So, we define the potential 
\be
\label{A=Abar+a}
A_a^\mu(\tau,x,y,\eta) \equiv \bar{A}_a^\mu(\tau,x,y,\eta)  + a_a^\mu(\tau,x,y,\eta),
\ee
such that $|\bar{A}_a^\mu(\tau,x,y,\eta)| \gg |a_a^\mu(\tau,x,y,\eta)|$. Substituting the potential (\ref{A=Abar+a}) into the Yang-Mills equations (\ref{YM-eq-tau-tau=0}) - (\ref{YM-eq-eta-tau=0}) and neglecting the terms non-linear in $a_a^\mu$, one finds 
\ba
\nn
&& - \, \partial_x\partial_\tau a^x_a 
- \partial_y  \partial_\tau a^y_a 
- \, \frac{1}{\tau^2} \, \partial_\eta \partial_\tau (\tau^2 a^\eta_a) 
\\[2mm]\label{YM-eq-tau-tau=0-lin}
&& ~~~~~~~~~~~~~
+ g \epsilon^{abc} \big(\bar{A}^x_b \partial_\tau a^x_c + a^x_b \partial_\tau \bar{A}^x_c 
+ \bar{A}^y_b \partial_\tau a^y_c + a^y_b \partial_\tau \bar{A}^y_c 
+ \bar{A}^\eta_b \partial_\tau (\tau^2 a^\eta_c) +
a^\eta_b \partial_\tau (\tau^2 \bar{A}^\eta_c) \big) = 0 ,
\ea
\ba
\nn
&&
 \partial_\tau^2 a^x_a 
+ \frac{1}{\tau} \, \partial_\tau a^x_a 
- \, \partial_y \big(\partial_y a^x_a - \partial_x a^y_a - g \epsilon^{abc}( \bar{A}^y_b a^x_c + a^y_b \bar{A}^x_c) \big)
- \frac{1}{\tau^2} \, \partial_\eta \big(\partial_\eta a^x_a - \tau^2 \partial_x a^\eta_a 
- g \tau^2 \epsilon^{abc} (\bar{A}^\eta_b a^x_c + a^\eta_b \bar{A}^x_c) \big)
\\[2mm] \nn
&& ~~~~~~~~~~~~~~~~
+ g \epsilon^{abc} \big( \bar{A}^y_b (\partial_y a^x_c - \partial_x a^y_c) 
+ a^y_b (\partial_y \bar{A}^x_c - \partial_x \bar{A}^y_c) 
+ \bar{A}^\eta_b (\partial_\eta a^x_c - \tau^2 \partial_x a^\eta_c) 
+ a^\eta_b (\partial_\eta \bar{A}^x_c - \tau^2 \partial_x \bar{A}^\eta_c) \big)
\\[2mm] \nn
&&~~~~~~~~~~~~~~~~~~~~~~~~~~
- g^2\big( \bar{A}^y_b \bar{A}^y_a a^x_b + \bar{A}^y_b a^y_a \bar{A}^x_b + a^y_b \bar{A}^y_a \bar{A}^x_b 
-\bar{A}^y_b \bar{A}^y_b a^x_a - \bar{A}^y_b a^y_b \bar{A}^x_a  - a^y_b \bar{A}^y_b \bar{A}^x_a 
\\[2mm] 
\label{YM-eq-x-tau=0-lin}
&&~~~~~~~~~~~~~~~~~~~~~~~~~~~~~~~~~~~~~~
+ \tau^2( \bar{A}^\eta_b \bar{A}^\eta_a a^x_b + \bar{A}^\eta_b a^\eta_a \bar{A}^x_b 
+ a^\eta_b \bar{A}^\eta_a \bar{A}^x_b 
-  \bar{A}^\eta_b \bar{A}^\eta_b a^x_a -  \bar{A}^\eta_b A^\eta_b \bar{A}^x_a
-  a^\eta_b \bar{A}^\eta_b \bar{A}^x_a )\big) 
= 0,
\ea
\ba
\nn
&& \partial_\tau^2 a^y_a  + \frac{1}{\tau} \, \partial_\tau a^y_a 
- \, \partial_x \big(\partial_x a^y_a - \partial_y a^x_a - g \epsilon^{abc}(\bar{A}^x_b a^y_c + a^x_b \bar{A}^y_c)\big) 
- \frac{1}{\tau^2} \, \partial_\eta \big(\partial_\eta a^y_a - \tau^2 \partial_y a^\eta_a 
- g \tau^2  \epsilon^{abc} (\bar{A}^\eta_b a^y_c + a^\eta_b \bar{A}^y_c) \big)  
\\[2mm] \nn
&& ~~~~~~~~~~~~~~~~
+ g \epsilon^{abc} \big( \bar{A}^x_b (\partial_x a^y_c - \partial_y a^x_c) 
+ a^x_b (\partial_x \bar{A}^y_c - \partial_y \bar{A}^x_c) 
+  \bar{A}^\eta_b (\partial_\eta a^y_c - \tau^2 \partial_y a^\eta_c) 
+ a^\eta_b (\partial_\eta \bar{A}^y_c - \tau^2 \partial_y \bar{A}^\eta_c)  \big)
\\[2mm] \nn
&& ~~~~~~~~~~~~~~~~~~~~~~~~~~
- g^2 (\bar{A}^x_b \bar{A}^x_a a^y_b + \bar{A}^x_b a^x_a \bar{A}^y_b + a^x_b \bar{A}^x_a \bar{A}^y_b
- \bar{A}^x_b \bar{A}^x_b a^y_a - \bar{A}^x_b a^x_b \bar{A}^y_a - a^x_b \bar{A}^x_b \bar{A}^y_a
\\[2mm] 
\label{YM-eq-y-tau=0-lin}
&&~~~~~~~~~~~~~~~~~~~~~~~~~~~~~~~~~~~~~~
+ \tau^2 (\bar{A}^\eta_b \bar{A}^\eta_a a^y_b + \bar{A}^\eta_b a^\eta_a \bar{A}^y_b 
+ a^\eta_b \bar{A}^\eta_a \bar{A}^y_b
- \bar{A}^\eta_b \bar{A}^\eta_b a^y_a - \bar{A}^\eta_b a^\eta_b \bar{A}^y_a 
- a^\eta_b \bar{A}^\eta_b \bar{A}^y_a) \big) = 0 ,
\ea
\ba
\nn
&& \partial_\tau^2 (\tau^2 a^\eta_a)
- \frac{1}{\tau}\partial_\tau (\tau^2 a^\eta_a) 
- \, \partial_x \big(\tau^2 \partial_x a^\eta_a - \partial_\eta a^x_a 
- g \tau^2 \epsilon^{abc} (\bar{A}^x_b a^\eta_c + a^x_b \bar{A}^\eta_c) \big)  
- \, \partial_y \big(\tau^2 \partial_y a^\eta_a - \partial_\eta a^y_a 
- g \tau^2 \epsilon^{abc}(\bar{A}^y_b a^\eta_c + a^y_b \bar{A}^\eta_c) \big)  
\\[2mm] \nn
&& ~~~~~~~~~~~~~~
+ g \epsilon^{abc} \big( \bar{A}^x_b (\tau^2 \partial_x a^\eta_c - \partial_\eta a^x_c) 
+ a^x_b (\tau^2 \partial_x \bar{A}^\eta_c - \partial_\eta \bar{A}^x_c) 
+ \bar{A}^y_b (\tau^2 \partial_y a^\eta_c - \partial_\eta a^y_c) 
+ a^y_b (\tau^2 \partial_y \bar{A}^\eta_c - \partial_\eta \bar{A}^y_c) \big) 
\\[2mm] \nn
&& ~~~~~~~~~~~~~~~~~~~~~~~~~~~~~~~~~~~~~~
+ g^2 \tau^2 (- \bar{A}^x_b \bar{A}^x_a a^\eta_b - \bar{A}^x_b a^x_a \bar{A}^\eta_b 
- a^x_b \bar{A}^x_a \bar{A}^\eta_b 
+ \bar{A}^x_b \bar{A}^x_b a^\eta_a + \bar{A}^x_b a^x_b \bar{A}^\eta_a 
+ a^x_b \bar{A}^x_b \bar{A}^\eta_a
\\[2mm] \label{YM-eq-eta-tau=0-lin}
&& ~~~~~~~~~~~~~~~~~~~~~~~~~~~~~~~~~~~~~~~~~~~~~~~~
- \bar{A}^y_b \bar{A}^y_a a^\eta_b - \bar{A}^y_b a^y_a \bar{A}^\eta_b - a^y_b \bar{A}^y_a \bar{A}^\eta_b 
+ \bar{A}^y_b \bar{A}^y_b a^\eta_a + \bar{A}^y_b a^y_b \bar{A}^\eta_a + a^y_b \bar{A}^y_b \bar{A}^\eta_a)
= 0.
\ea

\section{Stability of magnetic field}
\label{sec-stab-B}

We consider here a stability of the magnetic field $B$ generated along the axis $z$ at the earliest phase of a heavy-ion collision. The field is given by the formula (\ref{Bz-initial-adj}) that is it is assumed to be space-time uniform. A validity of the assumption is discussed in Sec.~\ref{sec-discussion}. 

The field occurs due to the pre-collision potentials $\beta_{1a}^i$ and $\beta_{2a}^i$ which are independent of each other. The two potentials are parametrized by means of two parameters $\lambda$ and $B$ in the following way
\be
\label{pre-coll-pot-B}
\beta_{1a}^i = \delta^{ix} \delta^{a3} \frac{1}{\lambda}\sqrt{\frac{B}{g}},
~~~~~~~~~~~~~~~~~
\beta_{2a}^i = \delta^{iy} \delta^{a2} \lambda \sqrt{\frac{B}{g}}.
\ee
The corresponding background four-potential can be written in the matrix notation as
\ba
\label{Abar-matrix-B}
\bar{A}^\mu_a = 
\left[ {\begin{array}{cccc}
0 ~~~~& 0 & 0 & 0 \\
0 ~~~~& 0 & \lambda \sqrt{\frac{B}{g}} & 0 \\
0 ~~~~& \frac{1}{\lambda} \sqrt{\frac{B}{g}} & 0 & 0 \\
 \end{array} } \right]  ,
\ea
where the Lorentz index $\mu$ numerates the columns and the color index $a$ numerates the rows. The potential (\ref{Abar-matrix-B}) substituted into Eqs.~(\ref{Ez-initial-adj}) and (\ref{Bz-initial-adj}) gives
\be
\label{Ez-Bz-fields}
E^\alpha_a = 0,
~~~~~~~~~~~~~~~~~
B^\alpha_a = \delta^{\alpha z} \delta^{a1} B. 
\ee
So, we have the vanishing electric field and uniform magnetic field of color 1 along the axis $z$. 

The magnetic field (\ref{Ez-Bz-fields}) is independent of $\lambda$ but the configurations of different lambdas are known to be physically nonequivalent \cite{Brown:1979bv}. It is easily demonstrated substituting the background potential (\ref{Abar-matrix-B}) into the Yang-Mills equations (\ref{YM-eq-tau-tau=0}) - (\ref{YM-eq-eta-tau=0}). Then, one finds
\ba
\label{j-matrix-B}
\left[ {\begin{array}{cccc}
0 ~~~~& 0 & 0 & 0 \\
0 ~~~~& 0 & \frac{1}{\lambda} \sqrt{gB^3} & 0 \\
0 ~~~~& \lambda \sqrt{g B^3} & 0 & 0 \\
\end{array} } \right]  
= j^\nu_a .
\ea
The current (\ref{j-matrix-B}) squared, which is gauge invariant, equals
\be
\label{jj}
j^\mu_a j_{a \,\mu} = - \Big(\frac{1}{\lambda^2} +\lambda^2\Big) g B^3 ,
\ee
and it depends on $(\lambda^{-2}+\lambda^2)$. Therefore, the potential configurations (\ref{Abar-matrix-B}) of different $(\lambda^{-2}+\lambda^2)$ are gauge nonequivalent \cite{Brown:1979bv}, even so they produce the same field strength and the same energy density. 

The equation (\ref{j-matrix-B}) also shows that, in contrast to the Abelian configuration of uniform magnetic field, the nonAbelian one (\ref{Abar-matrix-B}) does not solve the Yang-Mills equations with vanishing current. It means that the field must evolve in time to satisfy the equations. Consequently, a stationary character of the background filed is only an approximation which is applicable in the stability analysis if a rate of change of the background field is much smaller than a rate of change of small fluctuations. We return to this issue in Secs.~\ref{sec-glasma-evolution} and \ref{sec-discussion}. 

Performing the stability analysis we temporarily assume following \cite{Tudron:1980gq} that the current, which equals the left-hand side of Eq.~(\ref{j-matrix-B}), enters the Yang-Mills equation. Then, the stability analysis is performed in a standard way: the background potential (\ref{Abar-matrix-B}) solves the equations of motion and one checks whether small perturbations of this stationary solution grow or decay. 

With the background potential (\ref{Abar-matrix-B}), the linearized Yang-Mills equations (\ref{YM-eq-tau-tau=0-lin}) - (\ref{YM-eq-eta-tau=0-lin}) split into colors are
\ba
\label{YM-nA-B-tau-1}
&& - \, \partial_x\partial_\tau a^x_1 
- \partial_y  \partial_\tau a^y_1 
- \, \frac{1}{\tau^2} \, \partial_\eta \partial_\tau (\tau^2 a^\eta_1) 
+ g (- \bar{A}^x_3 \partial_\tau a^x_2 
+  \bar{A}^y_2 \partial_\tau a^y_3 ) = 0 ,
\\
\label{YM-nA-B-tau-2}
&& - \, \partial_x\partial_\tau a^x_2 
- \partial_y  \partial_\tau a^y_2 
- \, \frac{1}{\tau^2} \, \partial_\eta \partial_\tau (\tau^2 a^\eta_2) 
+ g \bar{A}^x_3 \partial_\tau a^x_1  = 0 ,
\\
\label{YM-nA-B-tau-3}
&& - \, \partial_x\partial_\tau a^x_3 
- \partial_y  \partial_\tau a^y_3 
- \, \frac{1}{\tau^2} \, \partial_\eta \partial_\tau (\tau^2 a^\eta_3) 
- g \bar{A}^y_2 \partial_\tau a^y_1 = 0 ,
\ea
\ba
\nn
&& \partial_\tau^2 a^x_1 
+ \frac{1}{\tau} \, \partial_\tau a^x_1 
- \, \partial_y \big(\partial_y a^x_1 - \partial_x a^y_1 - g (\bar{A}^y_2 a^x_3 + a^y_2 \bar{A}^x_3) \big)
- \frac{1}{\tau^2} \, \partial_\eta \big(\partial_\eta a^x_1 - \tau^2 \partial_x a^\eta_1 
- g \tau^2 a^\eta_2 \bar{A}^x_3 \big)
\\ \label{YM-nA-B-x-1}
&&~~~~~~~~~~~~~~~~~~~~~~~~~~~~~~~~~~~~~~~~~~~~~~~~~~~~~~~~~~~~~~~~~~~~~~~~~~~~
+ g\bar{A}^y_2 (\partial_y a^x_3 - \partial_x a^y_3) 
+ g^2 \bar{A}^y_2 \bar{A}^y_2 a^x_1  
= 0,
\\[2mm]
\label{YM-nA-B-x-2}
&& \partial_\tau^2 a^x_2 
+ \frac{1}{\tau} \, \partial_\tau a^x_2 
- \, \partial_y (\partial_y a^x_2 - \partial_x a^y_2 + g a^y_1 \bar{A}^x_3)
- \frac{1}{\tau^2} \, \partial_\eta (\partial_\eta a^x_2 - \tau^2 \partial_x a^\eta_2 
+ g \tau^2 a^\eta_1 \bar{A}^x_3 ) - g^2 a^y_3 \bar{A}^y_2 \bar{A}^x_3  = 0,
\\[2mm]
\nn
&&\partial_\tau^2 a^x_3 
+ \frac{1}{\tau} \, \partial_\tau a^x_3 
- \, \partial_y (\partial_y a^x_3 - \partial_x a^y_3 + g \bar{A}^y_2 a^x_1 )
- \frac{1}{\tau^2} \, \partial_\eta (\partial_\eta a^x_3 - \tau^2 \partial_x a^\eta_3 )
\\ \label{YM-nA-B-x-3}
&&~~~~~~~~~~~~~~~~~~~~~~~~~~~~~~~~~~~~~~~~~~~~~~~~~~~~~~~~~~~
- g \bar{A}^y_2 (\partial_y a^x_1 - \partial_x a^y_1) 
+ g^2(\bar{A}^y_2 \bar{A}^y_2 a^x_3 + 2 a^y_2 \bar{A}^y_2 \bar{A}^x_3 )
= 0,
\ea
\ba
\nn
&&\partial_\tau^2 a^y_1  + \frac{1}{\tau} \, \partial_\tau a^y_1 
- \, \partial_x \big(\partial_x a^y_1 - \partial_y a^x_1 + g (\bar{A}^x_3 a^y_2 + a^x_3 \bar{A}^y_2)\big) 
- \frac{1}{\tau^2} \, \partial_\eta \big(\partial_\eta a^y_1 
- \tau^2 \partial_y a^\eta_1 + g \tau^2  a^\eta_3 \bar{A}^y_2 \big)  
\\ \label{YM-nA-B-y-1}
&&~~~~~~~~~~~~~~~~~~~~~~~~~~~~~~~~~~~~~~~~~~~~~~~~~~~~~~~~~~~~~~~~~~~~~~~~~~~~~~
- g \bar{A}^x_3 (\partial_x a^y_2 - \partial_y a^x_2) 
+ g^2 \bar{A}^x_3 \bar{A}^x_3 a^y_1 = 0 ,
\\[2mm]
\nn
&& \partial_\tau^2 a^y_2  + \frac{1}{\tau} \, \partial_\tau a^y_2 
- \, \partial_x \big(\partial_x a^y_2 - \partial_y a^x_2 
- g \bar{A}^x_3 a^y_1) 
- \frac{1}{\tau^2} \, \partial_\eta (\partial_\eta a^y_2 - \tau^2 \partial_y a^\eta_2 )  
\\ \label{YM-nA-B-y-2}
&&~~~~~~~~~~~~~~~~~~~~~~~~~~~~~~~~~~~~~~~~~~~~~~~~~~~~~~~~~~
+ g \bar{A}^x_3 (\partial_x a^y_1 - \partial_y a^x_1) 
+ g^2 (\bar{A}^x_3 \bar{A}^x_3 a^y_2 + 2 a^x_3 \bar{A}^x_3 \bar{A}^y_2 ) = 0 ,
\\[2mm]
\label{YM-nA-B-y-3}
&& \partial_\tau^2 a^y_3  + \frac{1}{\tau} \, \partial_\tau a^y_3 
- \, \partial_x (\partial_x a^y_3 - \partial_y a^x_3 
- g a^x_1 \bar{A}^y_2)
- \frac{1}{\tau^2} \, \partial_\eta (\partial_\eta a^y_3 - \tau^2 \partial_y a^\eta_3 
- g \tau^2 a^\eta_1 \bar{A}^y_2 )  
- g^2 a^x_2 \bar{A}^x_3 \bar{A}^y_2 = 0 ,
\ea
\ba
\nn
&& \partial_\tau^2 (\tau^2 a^\eta_1)
- \frac{1}{\tau}\partial_\tau (\tau^2 a^\eta_1) 
- \, \partial_x (\tau^2 \partial_x a^\eta_1 - \partial_\eta a^x_1 
+ g \tau^2 \bar{A}^x_3 a^\eta_2 )  
- \, \partial_y (\tau^2 \partial_y a^\eta_1 - \partial_\eta a^y_1 
- g \tau^2 \bar{A}^y_2 a^\eta_3)  
\\  \label{YM-nA-B-eta-1}
&& ~~~~~~~~~~~~~~~~~~~~~~~~~~~~
- g  \big( \bar{A}^x_3 (\tau^2 \partial_x a^\eta_2 - \partial_\eta a^x_2) 
-  \bar{A}^y_2 (\tau^2 \partial_y a^\eta_3 - \partial_\eta a^y_3) \big) 
+ g^2 \tau^2 (\bar{A}^x_3 \bar{A}^x_3 + \bar{A}^y_2 \bar{A}^y_2 )  a^\eta_1 = 0,
\\[2mm]
\nn
&& \partial_\tau^2 (\tau^2 a^\eta_2)
- \frac{1}{\tau}\partial_\tau (\tau^2 a^\eta_2) 
- \, \partial_x (\tau^2 \partial_x a^\eta_2 - \partial_\eta a^x_2 
- g \tau^2  \bar{A}^x_3 a^\eta_1 )  
- \, \partial_y (\tau^2 \partial_y a^\eta_2 - \partial_\eta a^y_2) 
\\  \label{YM-nA-B-eta-2}
&& ~~~~~~~~~~~~~~~~~~~~~~~~~~~~~~~~~~~~~~~~~~~~~~~~~~~~~~~~~~~~~~~~~~~~~~~~~
+ g \bar{A}^x_3 (\tau^2 \partial_x a^\eta_1 - \partial_\eta a^x_1) 
+ g^2 \tau^2 \bar{A}^x_3 \bar{A}^x_3 a^\eta_2 = 0,
\\[2mm]
\nn
&& \partial_\tau^2 (\tau^2 a^\eta_3)
- \frac{1}{\tau}\partial_\tau (\tau^2 a^\eta_3) 
- \, \partial_x (\tau^2 \partial_x a^\eta_3 - \partial_\eta a^x_3) 
- \, \partial_y (\tau^2 \partial_y a^\eta_3 - \partial_\eta a^y_3 
+ g \tau^2 \bar{A}^y_2 a^\eta_1)  
\\ \label{YM-nA-B-eta-3}
&& ~~~~~~~~~~~~~~~~~~~~~~~~~~~~~~~~~~~~~~~~~~~~~~~~~~~~~~~~~~~~~~~~~~~~~~~~~~
- g  \bar{A}^y_2 (\tau^2 \partial_y a^\eta_1 - \partial_\eta a^y_1) 
+ g^2 \tau^2 \bar{A}^y_2 \bar{A}^y_2 a^\eta_3 = 0.
\ea

So, we have a system of 12 equations to be solved. We have managed to solve exactly the analogous set of equations using the Minkowski coordinates, background gauge and $\lambda = 1$ condition \cite{Bazak:2021xay}. Thanks to Minkowski coordinates the problem was fully algebraic after the Fourier transformation of space and time coordinates. Due to the background gauge a mixing of various color components was minimal, and the condition $\lambda = 1$ provided an axial symmetry with respect to the axis along the magnetic field. Here we deal with a more complicated system of equations. So, we consider a simplified situation when an evolution of longitudinal and transverse potential components is treated separately. Specifically, we discuss two special cases which still allow one to reveal characteristic features of the problem. 

\subsection{Special case: $a^\eta_a = 0 ~~ \&  ~~ a^x_a \not= 0,~ a^y_a \not= 0 $}
\label{sec-a-eta=0}

When $a^\eta_a = 0$, the equations (\ref{YM-nA-B-tau-1}) - (\ref{YM-nA-B-eta-3}) read
\ba
\label{YM-nA-B-tau-1-s1}
&& - \, \partial_x\partial_\tau a^x_1 
- \partial_y  \partial_\tau a^y_1 
- g (\bar{A}^x_3 \partial_\tau a^x_2 
- \bar{A}^y_2 \partial_\tau a^y_3 ) = 0 ,
\\
\label{YM-nA-B-tau-2-s1}
&& - \, \partial_x\partial_\tau a^x_2 
- \partial_y  \partial_\tau a^y_2 
+ g \bar{A}^x_3 \partial_\tau a^x_1  = 0 ,
\\
\label{YM-nA-B-tau-3-s1}
&& - \, \partial_x\partial_\tau a^x_3 
- \partial_y  \partial_\tau a^y_3 
- g \bar{A}^y_2 \partial_\tau a^y_1 = 0 ,
\ea
\ba
\nn
&&\Big(\partial_\tau^2 + \frac{1}{\tau} \, \partial_\tau - \partial_y^2 - \frac{1}{\tau^2} \, \partial_\eta^2 \Big) a^x_1 
+ \partial_x \partial_y a^y_1 
+ g \bar{A}^x_3 \partial_y a^y_2
\\ \label{YM-nA-B-x-1-s1}
&&~~~~~~~~~~~~~~~~~~~~~~~~~~~~~~~~~~~~
+ g\bar{A}^y_2 (2\partial_y a^x_3 - \partial_x a^y_3) 
+ g^2 \bar{A}^y_2 \bar{A}^y_2 a^x_1  
= 0,
\\[2mm]
\label{YM-nA-B-x-2-s1}
&& \Big(\partial_\tau^2 + \frac{1}{\tau} \, \partial_\tau - \partial_y^2 - \frac{1}{\tau^2} \, \partial_\eta^2 \Big) a^x_2 
+ \partial_x \partial_y a^y_2 - g \bar{A}^x_3 \partial_y a^y_1
- g^2 \bar{A}^y_2 \bar{A}^x_3  a^y_3 = 0,
\\[2mm]
\nn
&& \Big(\partial_\tau^2 + \frac{1}{\tau} \, \partial_\tau - \partial_y^2 - \frac{1}{\tau^2} \, \partial_\eta^2 \Big) a^x_3 
+ \partial_x \partial_y a^y_3  - g \bar{A}^y_2 (2\partial_y a^x_1 - \partial_x a^y_1) 
\\ \label{YM-nA-B-x-3-s1}
&&~~~~~~~~~~~~~~~~~~~~~~~~~~~~~~~~~~~~~~~~~~~~~~~
+ g^2(\bar{A}^y_2 \bar{A}^y_2 a^x_3 + 2 \bar{A}^y_2 \bar{A}^x_3 a^y_2)
= 0,
\ea
\ba
\nn
&&\Big(\partial_\tau^2 + \frac{1}{\tau} \, \partial_\tau - \partial_x^2 - \frac{1}{\tau^2} \, \partial_\eta^2\Big) a^y_1 
 + \partial_y \partial_x a^x_1 
- g \bar{A}^y_2 \partial_x a^x_3 
\\ \label{YM-nA-B-y-1-s1}
&&~~~~~~~~~~~~~~~~~~~~~~~~~~~~~~~~~~~~~
- g \bar{A}^x_3 (2\partial_x a^y_2 - \partial_y a^x_2) 
+ g^2 \bar{A}^x_3 \bar{A}^x_3 a^y_1 = 0 ,
\\[2mm]
\nn
&& \Big(\partial_\tau^2 + \frac{1}{\tau} \, \partial_\tau - \partial_x^2 - \frac{1}{\tau^2} \, \partial_\eta^2\Big) a^y_2  + \partial_y \partial_x a^x_2 
+ g \bar{A}^x_3 (2\partial_x a^y_1 - \partial_y a^x_1) 
\\ \label{YM-nA-B-y-2-s1}
&&~~~~~~~~~~~~~~~~~~~~~~~~~~~~~~~~~~~~~~~~~~~~~~~
+ g^2 (\bar{A}^x_3 \bar{A}^x_3 a^y_2 + 2 \bar{A}^x_3 \bar{A}^y_2 a^x_3) = 0 ,
\\[2mm]
\label{YM-nA-B-y-3-s1}
&& \Big(\partial_\tau^2 + \frac{1}{\tau} \, \partial_\tau - \partial_x^2 - \frac{1}{\tau^2} \, \partial_\eta^2\Big) a^y_3  + \partial_y \partial_x a^x_3 
+ g \bar{A}^y_2 \partial_x a^x_1
- g^2 \bar{A}^x_3 \bar{A}^y_2 a^x_2 = 0 ,
\ea
\ba
\label{YM-nA-B-eta-1-s1}
&& \partial_x \partial_\eta a^x_1 
+ \partial_y  \partial_\eta a^y_1 
+ g  ( \bar{A}^x_3 \partial_\eta a^x_2 
-  \bar{A}^y_2 \partial_\eta a^y_3)  = 0,
\\[2mm]
\label{YM-nA-B-eta-2-s1}
&& \partial_x \partial_\eta a^x_2 
+ \partial_y \partial_\eta a^y_2  
- g \bar{A}^x_3 \partial_\eta a^x_1 = 0,
\\[2mm]
\label{YM-nA-B-eta-3-s1}
&& \partial_x \partial_\eta a^x_3
+ \partial_y \partial_\eta a^y_3 
+ g  \bar{A}^y_2 \partial_\eta a^y_1 = 0.
\ea

One sees that the equations (\ref{YM-nA-B-tau-1-s1}) - (\ref{YM-nA-B-tau-3-s1}) and (\ref{YM-nA-B-eta-1-s1}) - (\ref{YM-nA-B-eta-3-s1}) are solved if
\ba
\label{YM-nA-B-eta-1-s1-sol}
&& \partial_x a^x_1 
+ \partial_y  a^y_1 
+ g  ( \bar{A}^x_3 a^x_2 
-  \bar{A}^y_2 a^y_3)  = 0,
\\[2mm]
\label{YM-nA-B-eta-2-s1-sol}
&& \partial_x a^x_2 
+ \partial_y a^y_2  
- g \bar{A}^x_3 a^x_1 = 0,
\\[2mm]
\label{YM-nA-B-eta-3-s1-sol}
&& \partial_x a^x_3
+ \partial_y  a^y_3 
+ g  \bar{A}^y_2 a^y_1 = 0.
\ea

Substituting $\partial_y  a^y_1 = - \partial_x a^x_1 - g  (\bar{A}^x_3 a^x_2 -  \bar{A}^y_2 a^y_3)$ and $\partial_y a^y_2 = - \partial_x a^x_2  + g \bar{A}^x_3 a^x_1$ into Eqs.~(\ref{YM-nA-B-x-1-s1}) and (\ref{YM-nA-B-x-2-s1}), and $\partial_y  a^y_3 = -\partial_x a^x_3 - g  \bar{A}^y_2 a^y_1$  into Eq.~(\ref{YM-nA-B-x-3-s1}), one finds
\ba
\label{YM-nA-B-x-1-s1-3}
&&\Big(\Box + g^2 (\bar{A}^x_3 \bar{A}^x_3 
+ \bar{A}^y_2 \bar{A}^y_2) \Big) a^x_1 
- 2 g \bar{A}^x_3 \partial_x a^x_2  
+ 2 g\bar{A}^y_2 \partial_y a^x_3 
= 0,
\\[2mm]
\label{YM-nA-B-x-2-s1-2}
&& \Big(\Box + g^2 \bar{A}^x_3 \bar{A}^x_3\Big) a^x_2 
+ 2g \bar{A}^x_3 \partial_x a^x_1
- 2g^2 \bar{A}^x_3 \bar{A}^y_2  a^y_3 
= 0,
\\[2mm]
 \label{YM-nA-B-x-3-s1-2}
&& \Big(\Box + g^2\bar{A}^y_2 \bar{A}^y_2 \Big) a^x_3 
- 2 g \bar{A}^y_2 \partial_y a^x_1 
+2 g^2 \bar{A}^y_2 \bar{A}^x_3 a^y_2
= 0,
\ea
where
\be
\Box \equiv \partial_\tau^2  + \frac{1}{\tau}\partial_\tau
- \partial_x^2 - \partial_y^2
- \frac{1}{\tau^2} \,\partial_\eta^2 .
\ee
Proceeding analogously with Eqs.~(\ref{YM-nA-B-y-1-s1}), (\ref{YM-nA-B-y-2-s1}) and (\ref{YM-nA-B-y-3-s1}), we obtain
\ba
\label{YM-nA-B-y-1-s1-3}
&&\Big( \Box + g^2 (\bar{A}^x_3 \bar{A}^x_3 + \bar{A}^y_2 \bar{A}^y_2)\Big) a^y_1 
+ 2g \bar{A}^y_2  \partial_y a^y_3
- 2g \bar{A}^x_3 \partial_x a^y_2
= 0 ,
\\[2mm]
\label{YM-nA-B-y-2-s1-3}
&& \Big( \Box + g^2 \bar{A}^x_3 \bar{A}^x_3 \Big) a^y_2  
+ 2g \bar{A}^x_3 \partial_x a^y_1 
+ 2 g^2  \bar{A}^x_3 \bar{A}^y_2 a^x_3 = 0 ,
\\[2mm]
\label{YM-nA-B-y-3-s1-3}
&& \Big( \Box + g^2 \bar{A}^y_2 \bar{A}^y_2 
\Big) a^y_3  - 2 g  \bar{A}^y_2  \partial_y a^y_1  
- 2 g^2 \bar{A}^y_2  \bar{A}^x_3 a^x_2 = 0 .
\ea

Defining the functions
\be
V^\pm_a \equiv a^x_a \pm ia^y_a ,
\ee
Eqs.~(\ref{YM-nA-B-x-1-s1-3}) - (\ref{YM-nA-B-y-3-s1-3}) provide
\ba
\label{V1-eq}
&&\Big(\Box + g^2 (\bar{A}^x_3 \bar{A}^x_3 
+ \bar{A}^y_2 \bar{A}^y_2) \Big) V^\pm_1 
- 2 g \bar{A}^x_3 \partial_x V^\pm_2  
+ 2 g\bar{A}^y_2 \partial_y V^\pm_3
= 0,
\\[2mm]
\label{V2-eq}
&& \Big(\Box + g^2 \bar{A}^x_3 \bar{A}^x_3\Big) V^\pm_2 
+ 2g \bar{A}^x_3 \partial_x V^\pm_1
\pm 2ig^2 \bar{A}^x_3 \bar{A}^y_2  V^\pm_3 
= 0,
\\[2mm]
 \label{V3-eq}
&& \Big(\Box + g^2\bar{A}^y_2 \bar{A}^y_2 \Big) V^\pm_3 
- 2 g \bar{A}^y_2 \partial_y V^\pm_1
\mp 2i g^2 \bar{A}^y_2 \bar{A}^x_3 V^\pm_2
= 0.
\ea
One sees that the equations of $V^+_1, V^+_2, V^+_3$ and those of $V^-_1, V^-_2, V^-_3$ form closed systems that is the functions $V^+_a$ do not mix up with $V^-_a$. 

Keeping in mind that $\bar{A}^y_2 = \lambda \sqrt{B/g}$ and $\bar{A}^x_3 = \lambda^{-1}\sqrt{B/g}$, Eqs.~(\ref{V1-eq}), (\ref{V2-eq}) and (\ref{V3-eq}) can be rewritten in the matrix notation as
\ba
\label{matrix-V-k-new}
\left[ \begin{array}{ccc}
\Box + (\lambda^2 + \lambda^{-2})gB & - 2 i\lambda^{-1} \sqrt{g B} \, k_x &   2i\lambda \sqrt{g B} \, k_y
\\ [2mm]
 2i \lambda^{-1} \sqrt{g B} \, k_x & \Box + \lambda^{-2}gB & \pm 2ig B  
\\ [2mm]
- 2i \lambda \sqrt{g B} \, k_y & \mp 2i g B & \Box + \lambda^2 gB
 \end{array}  \right]  
\left[ \begin{array}{c}
V^\pm_1 
\\ [2mm]
V^\pm_2 
\\ [2mm]
V^\pm_3
\end{array}  \right] 
= \left[ \begin{array}{c}
0
\\ [2mm]
0
\\ [2mm]
0
\end{array}  \right] ,
\ea
where we have assumed that the functions $V^\pm_a$ depend on $x,y,\eta$ through $e^{i(k_x x + k_y y + \nu \eta)}$ and consequently the d'Alembertian is redefined as
\be
\Box \equiv \partial_\tau^2  + \frac{1}{\tau}\partial_\tau +k_x^2 + k_y^2 + \frac{\nu^2}{\tau^2} .
\ee

The eigenvalues $\Lambda$ of the matrix in Eq.~(\ref{matrix-V-k-new}) are provided by the cubic equation
\ba
\label{det-matrix-Lambda}
{\rm det}\left[ \begin{array}{ccc}
\Box + (\lambda^2 + \lambda^{-2})gB - \Lambda 
& - 2 i\lambda^{-1} \sqrt{g B} \, k_x 
&   2i\lambda \sqrt{g B} \, k_y
\\ [2mm]
 2i\lambda^{-1} \sqrt{g B} \, k_x 
 & \Box + \lambda^{-2}gB - \Lambda 
 & \pm 2ig B  
\\ [2mm]
- 2i \lambda \sqrt{g B} \, k_y 
& \mp 2i g B 
& \Box + \lambda^2 gB - \Lambda
 \end{array}  \right] = 0 ,
\ea
which gives 
\ba
\nn
&& \big( \Box + (\lambda^2 + \lambda^{-2})gB - \Lambda \big)
\big( \Box + \lambda^{-2} gB - \Lambda \big)
\big( \Box + \lambda^2 gB - \Lambda \big)
\\[2mm] \label{eq-Lambda}
&&~~~~~~
-4g^2 B^2 \big( \Box + (\lambda^2 + \lambda^{-2})gB - \Lambda \big)
- 4 \lambda^{-2} g B \, k_x^2  \big( \Box + \lambda^2 gB - \Lambda \big)
- 4 \lambda^2 g B \, k_y^2 \big( \Box + \lambda^{-2} gB - \Lambda \big)
= 0.
\ea
One sees that the equation (\ref{eq-Lambda}) is the same for the equation of $V^+$ and $V^-$. Consequently, the eigenvalues of the matrices, which enter equations of $V^+$ and $V^-$, are the same. 

When $k_x=0$ and $k_y=0$, Eq.~(\ref{eq-Lambda}) becomes
\ba
\label{eq-Lambda-kx=ky=0}
\big( \Box + (\lambda^2 + \lambda^{-2})gB - \Lambda^{(0)} \big)
\Big(\Lambda^2  - \big( 2\Box + (\lambda^{-2} + \lambda^2) gB \big)\Lambda^{(0)}  
+ ( \Box + \lambda^{-2} gB)( \Box + \lambda^2 gB) - 4g^2 B^2 \Big)
= 0,
\ea
and it is solved by
\be
\label{0th-solutions-general}
\Lambda_1 = \Box + (\lambda^2 + \lambda^{-2})gB ,
~~~~~~~~~~~~~~~~~~~~~~
\Lambda_\pm = \Box + \frac{1}{2}\Big(\lambda^{-2} + \lambda^2 
\pm \sqrt{( \lambda^{-2} + \lambda^2)^2 + 12}\Big) gB .
\ee

After the diagonalization of the matrix equation (\ref{matrix-V-k-new}), one gets three equations  
\ba
 \label{f-1-k=0}
\Big(\partial_\tau^2  + \frac{1}{\tau}\partial_\tau + \frac{\nu^2}{\tau^2} + b_1^2\Big) f_1(\tau) &=& 0 ,
\\[2mm] \label{f-pm-k=0}
\Big(\partial_\tau^2  + \frac{1}{\tau}\partial_\tau + \frac{\nu^2}{\tau^2} \pm b_\pm^2 \Big) f_\pm(\tau) &=& 0 ,
\ea
where 
\ba
b_1^2 &\equiv& (\lambda^2 + \lambda^{-2})gB ,
\\[2mm]
b_\pm^2 &\equiv&  \frac{1}{2}\Big(\sqrt{( \lambda^{-2} + \lambda^2)^2 + 12} \,
\pm(\lambda^{-2} + \lambda^2)\Big) gB, 
\ea
and the functions $f_1$ and $f_\pm$ are linear combinations of either $V^+_1, V^+_2, V^+_3$ or $V^-_1, V^-_2, V^-_3$. When $\lambda=1$ we have
\be
b_1^2 = 2gB, ~~~~~~~~~~
b_+^2 = 3gB,~~~~~~~~~~
b_-^2 = gB .
\ee

One sees that the equations of $f_1$ and $f_+$ are the Bessel equation and that of $f_-$ is the modified Bessel equation. The equations are briefly discussed for a reader's convenience in Appendix. Since we are interested in solutions, which are everywhere finite, as we study small fluctuations around the background field, the solutions $f_1(\tau)$ and $f_+(\tau)$ are the Bessel function of imaginary order $J_{i\nu}(b_1 \tau)$ or $J_{i\nu}(b_+ \tau)$ and $f_-(\tau)$ is  the modified Bessel function $I_{i\nu}(b_- \tau)$. While the solutions $f_1(\tau)$ and $f_+(\tau)$ are oscillatory, the solution $f_-(\tau)$ grows as $\exp(\sqrt{b_- g B}\, \tau)$ when $\tau^2 > \nu^2/b_-^2$ and represents the instability analogous to the Nielsen-Olesen instability \cite{Nielsen:1978rm}. The difference is that the unstable Nielsen-Olesen mode appears in the Abelian configuration of the magnetic background field. We note that when $\lambda=1$ the equation (\ref{f-pm-k=0}) of $f_-$ fully coincides with that of the unstable mode of the Abelian configuration after a dependence on transverse coordinates is separated out using the Hamiltonian of harmonic oscillator \cite{Fujii:2008dd}. The time dependence of the unstable solution of Abelian configuration at any $k_x$ and $k_y$ coincides with that of the nonAbelian one at  $k_x=k_y=0$. One also checks that when $\lambda=1$ the growth rate of the unstable mode $\sqrt{b_- g B}$ is maximal and equal to $\sqrt{g B}$. When $\lambda$ goes to zero or infinity the growth rate tends to zero. 

It follows from the equation (\ref{f-pm-k=0}) that a behavior of the solution $f_-(\tau)$ depends of the sign of $\nu^2/\tau^2 - b_-^2$. For short times when $\nu^2/\tau^2 - b_-^2$ is positive the function $f_-(\tau)$ oscillates around zero, and for later times when $\nu^2/\tau^2 - b_-^2$ is negative the function exponentially grows, as discussed in \cite{Fukushima:2007yk,Iwazaki:2008xi,Fujii:2008dd}. When the field fluctuation is independent of space-time rapidity $\eta$, it is invariant under boosts along the axis $z$ and $\nu =0$. Then, the function $f_-(\tau)$ starts exponentially growing right from $\tau =0$. If the field fluctuation varies with $\eta$ and $\nu >0$, the exponential growth is delayed to $\tau = \nu/b_-$. 

When the stability analysis of uniform chromomagnetic field is performed using the Minkowski coordinates with time $t$ not the proper time $\tau$ \cite{Bazak:2021xay,Bazak:2022khg}, plane waves are solutions of the linearized equations of motion and the unstable modes start exponentially growing at $t=0$. So, there is no delay of the growth but the growth rate is reduced from $\sqrt{gB}$  to $\sqrt{gB - k_z^2}$ where $k_z$ is the longitudinal momentum analogous to $\nu$. 

\subsection{Effect of transverse momenta}
\label{sec-trans-momentum}

Let us now discuss how finite momenta $k_x$ or $k_y$ influence a stability of the uniform magnetic field. If $k_x \not= 0$ or $k_y \not= 0$,  the equation (\ref{eq-Lambda}) is a cubic one. Although roots of such an equation are given by the Cardano formula, their structure is rather complex. So, for finite $k_x$ or $k_y$ we solve Eq.~(\ref{eq-Lambda}) perturbatively, assuming that $gB \gg k_x^2$ and $gB \gg k_y^2$. The solutions (\ref{0th-solutions-general}) are now the zeroth order solution denoted as $\Lambda^{(0)}$, while the first order solutions are assumed to be of the form
\be
\label{1st-solution-general}
\Lambda^{(1)} = \Lambda^{(0)} + c_x k_x^2 + c_y k_y^2,
\ee
where the coefficients $c_x$ and $c_y$ are to be found. Substituting the formula (\ref{1st-solution-general}) into Eq.~(\ref{eq-Lambda}), neglecting the terms which are quadratic and cubic in $k_x^2$ and $k_y^2$, and using the zeroth order equations satisfied by  $\Lambda^{(0)}$, one finds
\ba
\nn
&& - \big(c_x k_x^2 + c_y k_y^2 \big) \Big[
\big( \Box + \lambda^{-2} gB - \Lambda^{(0)} \big)
\big( \Box + \lambda^2 gB - \Lambda^{(0)}  \big)
\\[2mm] \nn
&&~~~~~~~~~~~~~~~~~
+\big( \Box + (\lambda^2 + \lambda^{-2})gB -  \Lambda^{(0)}\big)
\big( \Box + \lambda^2 gB - \Lambda^{(0)}  \big)
\\[2mm] \nn
&&~~~~~~~~~~~~~~~~~
+\big( \Box + (\lambda^2 + \lambda^{-2})gB - \Lambda^{(0)}\big)
\big( \Box + \lambda^{-2} gB - \Lambda^{(0)} \big) - 4g^2 B^2 \Big]
\\[2mm] 
\label{eq-Lambda-1st-order}
&&~~~~~~~~~~~~~~~~~~~~~~~~~~~~~~
= 4 \lambda^{-2} g B \, k_x^2  \big( \Box + \lambda^2 gB - \Lambda^{(0)}\big)
+ 4 \lambda^2 g B \, k_y^2 \big( \Box + \lambda^{-2} gB - \Lambda^{(0)} \big) .
\ea
Since the equation (\ref{eq-Lambda-1st-order}) must be solved for any $k_x^2$ and any $k_y^2$, one finds the coefficients $c_x$ and $c_y$ putting  $k_y^2=0$ and $k_x^2=0$, respectively. Thus, one obtains
\ba
c_x &=& - \frac{4 \lambda^{-2}  \big(\lambda^2 - \bar\Lambda^{(0)}\big)}
{\big(\lambda^{-2} - \bar\Lambda^{(0)} \big)
\big(\lambda^2 - \bar\Lambda^{(0)}  \big)
+\big(\lambda^2 + \lambda^{-2} -  \bar\Lambda^{(0)}\big)
\big( \lambda^2 - \bar\Lambda^{(0)}  \big)
+\big(\lambda^2 + \lambda^{-2} - \bar\Lambda^{(0)}\big)
\big( \lambda^{-2} - \bar\Lambda^{(0)} \big) - 4 } ,
\\[2mm]
c_y &=& - \frac{4 \lambda^2  \big(\lambda^{-2} - \bar\Lambda^{(0)}\big)}
{\big(\lambda^{-2} - \bar\Lambda^{(0)} \big)
\big(\lambda^2 - \bar\Lambda^{(0)}  \big)
+\big(\lambda^2 + \lambda^{-2} -  \bar\Lambda^{(0)}\big)
\big( \lambda^2 - \bar\Lambda^{(0)}  \big)
+\big(\lambda^2 + \lambda^{-2} - \bar\Lambda^{(0)}\big)
\big( \lambda^{-2} - \bar\Lambda^{(0)} \big) - 4 } ,
\ea
where 
\be
\bar\Lambda^{(0)} \equiv \frac{\Lambda^{(0)} - \Box}{gB} .
\ee

Using the zeroth order solutions $\Lambda^{(0)}_1$ and $\Lambda^{(0)}_\pm$, we obtain
\ba
\label{cxy1}
c_x^1 &=& = - \frac{4}{3}  \lambda^{-4}, ~~~~~~~~~
c_y^1  = - \frac{4}{3}  \lambda^4 ,
\\[2mm]
\label{cxpm}
c_x^\pm &=& - \frac{4 \lambda^{-2}  \Big(-\lambda^{-2} + \lambda^2 
\mp \sqrt{( \lambda^{-2} + \lambda^2)^2 + 12}\Big)}
{\lambda^4 + \lambda^{-4}  
\mp (\lambda^2 + \lambda^{-2}) \sqrt{( \lambda^{-2} + \lambda^2)^2 + 12} + 14} ,
\\[2mm]
\label{cypm}
c_y^\pm  &=& - \frac{4 \lambda^2 \Big(\lambda^{-2} - \lambda^2 
\mp \sqrt{( \lambda^{-2} + \lambda^2)^2 + 12}\Big)}
{\lambda^4 + \lambda^{-4}  
\mp (\lambda^2 + \lambda^{-2}) \sqrt{( \lambda^{-2} + \lambda^2)^2 + 12} + 14} .
\ea
For $\lambda =1$ the formulas (\ref{cxy1}), (\ref{cxpm}) and (\ref{cypm}) simplify to
\be
c_x^1 = c_y^1 =  - \frac{4}{3}, 
~~~~~~~~~~~~~~~~
c_x^\pm  = c_y^\pm  = \pm \frac{2}{2 \mp 1} .
\ee

Knowing the approximate eigenvalues of the matrix from Eq.~(\ref{matrix-V-k-new}), we can write down the equations of motion as
\ba
\label{f-1-k-not=0}
&& \Big(\partial_\tau^2  + \frac{1}{\tau}\partial_\tau + \frac{\nu^2}{\tau^2} 
+ b_1^2 + (c_x^1 +1) k_x^2 + (c_y^1 +1) k_y^2  \Big) f_1(\tau) = 0,
\\[2mm] 
\label{f-pm-k-not=0}
&& \Big(\partial_\tau^2  + \frac{1}{\tau}\partial_\tau  + \frac{\nu^2}{\tau^2} 
\pm b_\pm^2 + (c_x^\pm +1) k_x^2 + (c_y^\pm +1) k_y^2  \Big) f_\pm(\tau) = 0.
\ea

Since $k_x^2 \ll gB$ and $k_y^2 \ll gB$, a character of the solutions of Eqs.~(\ref{f-1-k-not=0}) and (\ref{f-pm-k-not=0}) is similar to those of Eqs.~(\ref{f-1-k=0}) and (\ref{f-pm-k=0}). The solutions $f_1$ and $f_+$ are stable while $f_-$ can be unstable. So, we focus on the latter one which is of our main interest. The solution is unstable if $\nu^2/\tau^2 
- b_-^2 + (c_x^- +1) k_x^2 + (c_y^- +1) k_y^2$ is negative. Since $c_x^- +1$ and $c_y^- +1$ are both positive for any $\lambda$, one finds that finite momenta $k_x$ and $k_y$ tend to stabilize the solution. It is clearly seen in case of $\lambda=1$ when, as already discussed, the growth rate  $\sqrt{b_- g B}$ is maximal when $k_x=k_y=0$. For $\lambda=1$ the equation of $f_-$  reads
\be
 \Big(\partial_\tau^2  + \frac{1}{\tau}\partial_\tau  + \frac{\nu^2}{\tau^2} 
- gB + \frac{1}{3}k_T^2 \Big) f_-(\tau) = 0,
\ee
where $k_T^2 \equiv k_x^2 + k_y^2$. The growth rate of the unstable mode is $\sqrt{g B - k_T^2/3}$. However, we cannot conclude that the instability is absent if $k_T^2 \ge 3 gB$, as $k_T^2$ is assumed to be small when compared to $gB$. We note that in case of the Nielsen-Olesen mode in the Abelian configuration of chromomagnetic field the growth rate is not influenced by the transverse momentum and it equals $\sqrt{g B}$. A comparison of unstable modes in Abelian and nonAbelian configurations of uniform chromomagnetic field is presented in Fig.~5 of our eariler work \cite{Bazak:2021xay} where the stability analysis is performed using the Minkowski coordinates with time not the proper time. 

\subsection{Special case: $a^\eta_a \not= 0 ~~ \&  ~~ a^x_a = a^y_a = 0$}

Let us now discuss purely longitudinal dynamics. When $a^x_a = a^y_a = 0$, the equations (\ref{YM-nA-B-tau-1}) - (\ref{YM-nA-B-eta-3}) read
\ba
\label{YM-nA-B-ax=ay=0-tau-1}
&&\partial_\eta \partial_\tau (\tau^2 a^\eta_1) = 0 ,
\\
\label{YM-nA-B-ax=ay=0-tau-2}
&&\partial_\eta \partial_\tau (\tau^2 a^\eta_2)  = 0 ,
\\
\label{YM-nA-B-ax=ay=0-tau-3}
&& \partial_\eta \partial_\tau (\tau^2 a^\eta_3)  = 0 ,
\ea
\ba
\label{YM-nA-ax=ay=0-B-x-1}
&&  \partial_\eta \big(\partial_x a^\eta_1 - g a^\eta_2 \bar{A}^x_3 \big)
= 0,
\\
\label{YM-nA-B-ax=ay=0-x-2}
&& \partial_\eta ( \partial_x a^\eta_2 
+ g a^\eta_1 \bar{A}^x_3 ) = 0,
\\
\label{YM-nA-B-ax=ay=0-x-3}
&&\partial_\eta \partial_x a^\eta_3 = 0,
\ea
\ba
\label{YM-nA-B-ax=ay=0-y-1}
&&
\partial_\eta \big(\partial_y a^\eta_1 - g  a^\eta_3 \bar{A}^y_2 \big)  = 0 ,
\\
\label{YM-nA-B-ax=ay=0-y-2}
&& \partial_\eta \partial_y a^\eta_2   = 0 ,
\\
\label{YM-nA-B-ax=ay=0-y-3}
&&  \partial_\eta (\partial_y a^\eta_3 + g a^\eta_1 \bar{A}^y_2 )  = 0 ,
\ea
\ba
\label{YM-nA-B-ax=ay=0-eta-1}
&& 
\Big(\partial_\tau^2  + \frac{1}{\tau}\partial_\tau  - \partial_x^2 - \partial_y^2
+ g^2 (\bar{A}^x_3 \bar{A}^x_3 + \bar{A}^y_2 \bar{A}^y_2 ) \Big) a^\eta_1
- 2g \bar{A}^x_3 \partial_x a^\eta_2  
+ 2g \bar{A}^y_2 \partial_y a^\eta_3  = 0,
\\
\label{YM-nA-B-ax=ay=0-eta-2}
&& 
\Big(\partial_\tau^2  + \frac{1}{\tau}\partial_\tau - \partial_x^2 -\partial_y^2
+ g^2 \bar{A}^x_3 \bar{A}^x_3 \Big) a^\eta_2
+  2g \bar{A}^x_3 \partial_x a^\eta_1 = 0,
\\
\label{YM-nA-B-ax=ay=0-eta-3}
&& 
\Big(\partial_\tau^2  + \frac{1}{\tau}\partial_\tau - \partial_x^2 -\partial_y^2 
+ g^2 \bar{A}^y_2 \bar{A}^y_2 \Big) a^\eta_3
- 2g  \bar{A}^y_2 \partial_y a^\eta_1  = 0.
\ea

One sees that Eqs.~(\ref{YM-nA-B-ax=ay=0-tau-1}) - (\ref{YM-nA-B-ax=ay=0-y-3}) are solved if $a^\eta_a$ is independent of $\eta$ which is the case assumed further on. Keeping in mind that $\bar{A}^y_2 = \lambda \sqrt{B/g}$ and $\bar{A}^x_3 = \lambda^{-1}\sqrt{B/g}$, Eqs.~(\ref{YM-nA-B-ax=ay=0-eta-1}), (\ref{YM-nA-B-ax=ay=0-eta-2}) and (\ref{YM-nA-B-ax=ay=0-eta-3}) can be rewritten in the matrix notation as
\ba
\label{matrix-a-eta-k}
\left[ \begin{array}{ccc}
\Box + (\lambda^2 + \lambda^{-2})gB & - 2 i\lambda^{-1} \sqrt{g B} \, k_x &   2i\lambda \sqrt{g B} \, k_y
\\ [2mm]
 2\lambda^{-1} i \sqrt{g B} \, k_x & \Box + \lambda^{-2}gB & 0
\\ [2mm]
- 2\lambda i \sqrt{g B} \, k_y &0 & \Box + \lambda^2 gB
 \end{array}  \right]  
\left[ \begin{array}{c}
a^\eta_1 
\\ [2mm]
a^\eta_2
\\ [2mm]
a^\eta_3
\end{array}  \right] 
= \left[ \begin{array}{c}
0
\\ [2mm]
0
\\ [2mm]
0
\end{array}  \right] ,
\ea
where we have assumed that the functions $a^\eta_a$ depend on $x$ and $y$ through $e^{i(k_x x + k_y y)}$ and consequently 
\be
\Box \equiv \partial_\tau^2  + \frac{1}{\tau}\partial_\tau +k_x^2 + k_y^2 .
\ee

When $k_x = k_y = 0$, Eq.~(\ref{matrix-a-eta-k}) is diagonal and it provides three equations
\ba
\label{YM-nA-B-ax=ay=0-k=0-eta-1}
&& 
\Big(\partial_\tau^2  + \frac{1}{\tau}\partial_\tau  
+ (\lambda^{-2}+\lambda^2)gB \Big) a^\eta_1  = 0,
\\
\label{YM-nA-B-ax=ay=0-k=0-eta-2}
&& 
\Big(\partial_\tau^2  + \frac{1}{\tau}\partial_\tau
+ \lambda^{-2} gB \Big) a^\eta_2 = 0,
\\
\label{YM-nA-B-ax=ay=0-k=0-eta-3}
&& 
\Big(\partial_\tau^2  + \frac{1}{\tau}\partial_\tau
+ \lambda^2gB \Big) a^\eta_3 = 0.
\ea
The solutions are $a^\eta_i \sim J_0(\sigma_i \tau)$ with 
\be
\sigma_1 = \sqrt{ (\lambda^{-2}+\lambda^2)gB },
~~~~~~~~~~~~~~~~
\sigma_2 = \lambda^{-1}\sqrt{gB },
~~~~~~~~~~~~~~~~
\sigma_3 = \lambda \sqrt{gB }.
\ee
The solutions are stable.

When $k_x \not=0$ or  $k_y \not= 0$, Eq.~(\ref{matrix-a-eta-k}) needs to be diagonalized. In case $\lambda =1$, one easily finds three equations
\ba
&&\Big(\partial_\tau^2  + \frac{1}{\tau}\partial_\tau + k_T^2 
+ gB\Big)f_1 = 0,
\\
&&\Big(\partial_\tau^2  + \frac{1}{\tau}\partial_\tau + k_T^2 
+ \frac{1}{2}\big(3gB \pm \sqrt{g^2B^2 + 16 gB k_T^2}\,\,\big) \Big)f_\pm = 0.
\ea
Because $k_T^2 + gB > 0$ and
$$
 k_T^2 + \frac{1}{2}\big(3gB \pm \sqrt{g^2B^2 + 16 gB k_T^2}\,\,\big) > 0 ,
$$
the solutions are stable. 

We have explicitly shown above that the uniform longitudinal chromomagnetic field is stable under purely longitudinal fluctuations in two special cases: 1) $\lambda$ arbitrary and $k_x =k_y = 0$, 2) $\lambda = 1$  and $k_x$, $ k_y$ are arbitrary. When $\lambda$, $k_x$ and $k_y$ are arbitrary, the situation is more complicated but the special cases suggest that the solutions of interest are stable.

\section{Stability of electric field}
\label{sec-stab-E}

We consider here a stability of the uniform electric field $E$ along the collision axis $z$. The field is in the Abelian configuration given by the formula (\ref{Ez-initial-adj}).  In Minkowski coordinates, the potential that generates the electric field $E^i_a = \delta^{iz}\delta^{a1} E$ and obeys the gauge condition $t\bar{A}^t - z \bar{A}^z = 0$ is 
\be
\bar{A}^\mu_a = - \frac{1}{2} \delta^{a1} (zE,0, 0, tE) .
\ee
In Milne coordinates, the potential is
\be
\label{background-pot-E}
\bar{A}^\mu_a = - \frac{1}{2} \delta^{a1} (0,0, 0, E) ,
~~~~~~~~~~~~~~~~
\bar{A}_\mu^a = \frac{1}{2} \delta^{a1} (0,0, 0, \tau^2 E) .
\ee

The problem of stability of uniform chromoelectric field in the Abelian configuration was studied in Minkowski coordinates using the axial gauge $\bar{A}^z=a^z=0$ in \cite{Chang:1979tg} and the Lorentz gauge $\partial_\mu \bar{A}^\mu =0$  combined with the background gauge $D_\mu a^\mu =0$, where the covariant derivative $D_\mu$ includes the background potential, in our earlier work \cite{Bazak:2021xay}. The problem was also studied in the context of glasma \cite{Fujii:2008dd}, using the Milne coordinates and Fock-Schwinger gauge (\ref{gauge-condition}). In our analysis, which is presented below, we follow the study \cite{Fujii:2008dd}, clarifying and improving some points. 

With the background potential (\ref{background-pot-E}), the linearized Yang-Mills equations  (\ref{YM-eq-tau-tau=0-lin}) - (\ref{YM-eq-eta-tau=0-lin}) split into color components are
\ba
\label{YM-eq-tau-E-1}
&& - \, \partial_x\partial_\tau a^x_1 
- \partial_y  \partial_\tau a^y_1 
- \, \frac{1}{\tau^2} \, \partial_\eta \partial_\tau (\tau^2 a^\eta_1)  = 0 ,
\\[2mm]
\label{YM-eq-tau-E-2}
&& - \, \partial_x\partial_\tau a^x_2 
- \partial_y  \partial_\tau a^y_2 
- \, \frac{1}{\tau^2} \, \partial_\eta \partial_\tau (\tau^2 a^\eta_2) 
- g \tau^2 \bar{A}^\eta_1  \partial_\tau a^\eta_3 = 0 ,
\\[2mm]
\label{YM-eq-tau-E-3}
&& - \, \partial_x\partial_\tau a^x_3 
- \partial_y  \partial_\tau a^y_3 
- \, \frac{1}{\tau^2} \, \partial_\eta \partial_\tau (\tau^2 a^\eta_3) 
+ g \tau^2 \bar{A}^\eta_1 \partial_\tau a^\eta_2 = 0 ,
\ea
\ba
 \label{YM-eq-x-E-1}
&& \partial_\tau^2 a^x_1 
+ \frac{1}{\tau} \, \partial_\tau a^x_1 
- \, \partial_y \big(\partial_y a^x_1 - \partial_x a^y_1 \big)
- \frac{1}{\tau^2} \, \partial_\eta \big(\partial_\eta a^x_1 - \tau^2 \partial_x a^\eta_1  \big) = 0,
\\[2mm] \nn
&& \partial_\tau^2 a^x_2
+ \frac{1}{\tau} \, \partial_\tau a^x_2 
- \, \partial_y \big(\partial_y a^x_2 - \partial_x a^y_2 \big)
- \frac{1}{\tau^2} \, \partial_\eta \big(\partial_\eta a^x_2 - \tau^2 \partial_x a^\eta_2 
+ g \tau^2 \bar{A}^\eta_1 a^x_3 \big)
\\[2mm]  \label{YM-eq-x-E-2}
&& ~~~~~~~~~~~~~~~~~~~~~~~~~~~~~~~~~~~~~~~~~~
- g \bar{A}^\eta_1 (\partial_\eta a^x_3 - \tau^2 \partial_x a^\eta_3) 
+ g^2\tau^2 \bar{A}^\eta_1 \bar{A}^\eta_1 a^x_2 = 0,
\\[2mm] \nn
&& \partial_\tau^2 a^x_3 
+ \frac{1}{\tau} \, \partial_\tau a^x_3 
- \, \partial_y \big(\partial_y a^x_3 - \partial_x a^y_3 \big)
- \frac{1}{\tau^2} \, \partial_\eta \big(\partial_\eta a^x_3 - \tau^2 \partial_x a^\eta_3 
- g \tau^2 \bar{A}^\eta_1 a^x_2 \big)
\\[2mm]   \label{YM-eq-x-E-3}
&& ~~~~~~~~~~~~~~~~~~~~~~~~~~~~~~~~~~~~~~~~~~
+ g \bar{A}^\eta_1 (\partial_\eta a^x_2 - \tau^2 \partial_x a^\eta_2) 
+ g^2\tau^2 \bar{A}^\eta_1 \bar{A}^\eta_1 a^x_3  = 0,
\ea
\ba
\label{YM-eq-y-E-1}
&& \partial_\tau^2 a^y_1  + \frac{1}{\tau} \, \partial_\tau a^y_1 
- \, \partial_x \big(\partial_x a^y_1 - \partial_y a^x_1\big) 
- \frac{1}{\tau^2} \, \partial_\eta \big(\partial_\eta a^y_1 - \tau^2 \partial_y a^\eta_1 \big)  
 = 0 ,
\\[2mm]
\nn
&& \partial_\tau^2 a^y_2  + \frac{1}{\tau} \, \partial_\tau a^y_2 
- \, \partial_x \big(\partial_x a^y_2 - \partial_y a^x_2\big) 
- \frac{1}{\tau^2} \, \partial_\eta \big(\partial_\eta a^y_2 - \tau^2 \partial_y a^\eta_2 
+ g \tau^2  \bar{A}^\eta_1 a^y_3 \big)  
\\[2mm] \label{YM-eq-y-E-2}
&&~~~~~~~~~~~~~~~~~~~~~~~~~~~~~~~~~~~~~~~~~~
- g \bar{A}^\eta_1 (\partial_\eta a^y_3 - \tau^2 \partial_y a^\eta_3) 
+ g^2 \tau^2 \bar{A}^\eta_1 \bar{A}^\eta_1 a^y_2 = 0 ,
\\[2mm]
\nn
&& \partial_\tau^2 a^y_3  + \frac{1}{\tau} \, \partial_\tau a^y_3 
- \, \partial_x \big(\partial_x a^y_3 - \partial_y a^x_3\big) 
- \frac{1}{\tau^2} \, \partial_\eta \big(\partial_\eta a^y_3 - \tau^2 \partial_y a^\eta_3 
- g \tau^2 \bar{A}^\eta_1 a^y_2 \big)  
\\[2mm] \label{YM-eq-y-E-3}
&&~~~~~~~~~~~~~~~~~~~~~~~~~~~~~~~~~~~~~~~~~~
+ g \bar{A}^\eta_1 (\partial_\eta a^y_2 - \tau^2 \partial_y a^\eta_2) 
+ g^2 \tau^2 \bar{A}^\eta_1 \bar{A}^\eta_1 a^y_3 = 0 ,
\ea
\ba
\label{YM-eq-eta-E-1}
&& \partial_\tau^2 (\tau^2 a^\eta_1)
- \frac{1}{\tau}\partial_\tau (\tau^2 a^\eta_1) 
- \, \partial_x \big(\tau^2 \partial_x a^\eta_1 - \partial_\eta a^x_1 \big)  
- \, \partial_y \big(\tau^2 \partial_y a^\eta_1 - \partial_\eta a^y_1 \big)  = 0,
\\[2mm]\nn
&& \partial_\tau^2 (\tau^2 a^\eta_2)
- \frac{1}{\tau}\partial_\tau (\tau^2 a^\eta_2) 
- \, \partial_x \big(\tau^2 \partial_x a^\eta_2 - \partial_\eta a^x_2 
- g \tau^2 a^x_3 \bar{A}^\eta_1 \big)  
\\[2mm] \label{YM-eq-eta-E-2}
&& ~~~~~~~~~~~~~~~~~~~~~~~~~~~~~~~~~~~~~~~~~~~
- \, \partial_y \big(\tau^2 \partial_y a^\eta_2 - \partial_\eta a^y_2 
- g \tau^2 a^y_3 \bar{A}^\eta_1 \big) = 0,
\\[2mm]\nn
&& \partial_\tau^2 (\tau^2 a^\eta_3)
- \frac{1}{\tau}\partial_\tau (\tau^2 a^\eta_3) 
- \, \partial_x \big(\tau^2 \partial_x a^\eta_3 - \partial_\eta a^x_3 
+ g \tau^2 a^x_2 \bar{A}^\eta_1 \big)  
\\[2mm] \label{YM-eq-eta-E-3}
&& ~~~~~~~~~~~~~~~~~~~~~~~~~~~~~~~~~~~~~~~~~~~~
- \, \partial_y \big(\tau^2 \partial_y a^\eta_3 - \partial_\eta a^y_3 
+ g \tau^2 a^y_2 \bar{A}^\eta_1 \big) = 0.
\ea

As one observes, the fluctuating potential $a^\mu_1$ is decoupled not only from $a^\mu_2$ and $a^\mu_3$ but from the background field as well. So, $a^\mu_1$ describes free waves which are not discussed any more. However, we still have a set of 8 equations. Following \cite{Fujii:2008dd}, we consider a simplified situation when an evolution of longitudinal and transverse potential components is treated separately. Specifically, we discuss two special cases which allow one to reveal characteristic features of the problem. 

\subsection{Special case: $a^x_a = a^y_a = 0 ~~ \&  ~~ a^\eta_a \not= 0$}

When $a^x_a = a^y_a = 0$ and $a^\eta_a \not= 0$, Eqs.~(\ref{YM-eq-tau-E-1}) - (\ref{YM-eq-eta-E-3}) are 
\ba
\label{YM-eq-tau-E-2-sc1}
&& \Big( \partial_\tau + \frac{2}{\tau} \Big) \partial_\eta a^\eta_2
+ g \tau^2 \bar{A}^\eta_1  \partial_\tau a^\eta_3 = 0 ,
\\[2mm]
\label{YM-eq-tau-E-3-sc1}
&&  \Big(\partial_\tau + \frac{2}{\tau} \Big) \partial_\eta a^\eta_3  
- g \tau^2 \bar{A}^\eta_1 \partial_\tau a^\eta_2 = 0 ,
\ea
\ba
\label{YM-eq-x-E-2-sc1}
&& \partial_x \Big( \partial_\eta a^\eta_2 
+g \tau^2 \bar{A}^\eta_1  a^\eta_3  \Big) = 0,
\\[2mm] 
 \label{YM-eq-x-E-3-sc1}
&&\partial_x \Big(\partial_\eta a^\eta_3 
- g \tau^2 \bar{A}^\eta_1  a^\eta_2  \Big) = 0,
\ea
\ba
\label{YM-eq-y-E-2-sc1}
&& \partial_y \Big(\partial_\eta a^\eta_2 
+ g \tau^2 \bar{A}^\eta_1 a^\eta_3 \Big) = 0 ,
\\[2mm]
\label{YM-eq-y-E-3-sc1}
&& \partial_y \Big( \partial_\eta a^\eta_3 
- g \tau^2 \bar{A}^\eta_1 a^\eta_2 \Big) = 0 ,
\ea
\ba
\label{YM-eq-eta-E-2-sc1}
&& \Big(\partial_\tau^2 + \frac{3}{\tau}\partial_\tau - \partial_x^2 - \partial_y^2 \Big) a^\eta_2  = 0,
\\[2mm]
\label{YM-eq-eta-E-3-sc1}
&& \Big(\partial_\tau^2 + \frac{3}{\tau}\partial_\tau - \partial_x^2 - \partial_y^2 \Big) a^\eta_3 = 0.
\ea

Introducing the functions 
\be
H^\pm(\tau, x, y, \eta) \equiv  a^\eta_2(\tau, x, y, \eta) \pm i a^\eta_3(\tau, x, y, \eta) ,
\ee
Eqs.~(\ref{YM-eq-tau-E-2-sc1}) - (\ref{YM-eq-eta-E-3-sc1}) are written as
\ba
\label{H-eq-E-tau-sc1}
&& \Big[\Big( \partial_\tau + \frac{2}{\tau} \Big) \partial_\eta 
\mp i g \tau^2 \bar{A}^\eta_1  \partial_\tau \Big] H^\pm = 0 ,
\\[2mm]
\label{H-eq-E-x-sc1}
&& \partial_x \Big( \partial_\eta  
\mp i g \tau^2 \bar{A}^\eta_1  \Big) H^\pm = 0,
\\[2mm] 
\label{H-eq-E-y-sc1}
&& \partial_y \Big( \partial_\eta  
\mp i g \tau^2 \bar{A}^\eta_1  \Big) H^\pm = 0,
\\[2mm]
\label{H-eq-E-eta-sc1}
&& \Big(\partial_\tau^2 + \frac{3}{\tau}\partial_\tau 
- \partial_x^2 - \partial_y^2 \Big) H^\pm   = 0 .
\ea
If $\partial_x H^\pm \not= 0$ and $\partial_y H^\pm \not= 0$, the equations (\ref{H-eq-E-x-sc1}) and (\ref{H-eq-E-y-sc1}) are solved if the functions $H^\pm$ obey
\be
\label{H-eq-777}
\Big( \partial_\eta  \mp i g \tau^2 \bar{A}^\eta_1  \Big) H^\pm = 0. 
\ee
Eq.~(\ref{H-eq-777}) is solved by $H^\pm \sim \exp(\pm ig \tau^2 \bar{A}^\eta_1 \,\eta)$ which substituted to into Eq~(\ref{H-eq-E-tau-sc1}) gives  $H^\pm =0$. So, we conclude that if  $a^x_a = a^y_a = 0$ then $ a^\eta_a = 0$ as well. In other words, the purely longitudinal dynamics is trivial. 

\subsection{Special case: $a^\eta_a = 0 ~~ \&  ~~ a^x_a \not= 0,~ a^y_a \not= 0 $}

When  $a^\eta_a = 0$ and $a^x_a \not= 0$ or $a^y_a \not= 0$, Eqs.~(\ref{YM-eq-tau-E-1}) - (\ref{YM-eq-eta-E-3}) read
\ba
\label{YM-eq-tau-E-2-sc2}
&& \partial_\tau \Big(\partial_x a^x_2 + \partial_y a^y_2  \Big) = 0 ,
\\[2mm]
\label{YM-eq-tau-E-3-sc2}
&&\partial_\tau \Big(\partial_x a^x_3 + \partial_y a^y_3  \Big) = 0 ,
\ea
\ba
 \label{YM-eq-x-E-2-sc2}
&&  \Big( \partial_\tau^2 + \frac{1}{\tau} \, \partial_\tau 
- \partial_y^2 - \frac{1}{\tau^2} \, \partial_\eta^2 + g^2\tau^2 \bar{A}^\eta_1 \bar{A}^\eta_1\Big) a^x_2
+ \partial_x \partial_y a^y_2 
- 2 g \bar{A}^\eta_1\partial_\eta a^x_3  = 0,
\\[2mm] 
\label{YM-eq-x-E-3-sc2}
&&  \Big( \partial_\tau^2  + \frac{1}{\tau} \, \partial_\tau - \partial_y^2 
- \frac{1}{\tau^2} \, \partial_\eta^2 
+ g^2\tau^2 \bar{A}^\eta_1 \bar{A}^\eta_1 \Big)  a^x_3
+ \partial_x \partial_y a^y_3 
+ 2g \bar{A}^\eta_1 \partial_\eta  a^x_2 = 0,
\ea
\ba
 \label{YM-eq-y-E-2-sc2}
&&  \Big( \partial_\tau^2  + \frac{1}{\tau} \, \partial_\tau - \partial_x^2 
- \frac{1}{\tau^2} \, \partial_\eta^2 
+ g^2 \tau^2 \bar{A}^\eta_1 \bar{A}^\eta_1 \Big) a^y_2 
+  \partial_y \partial_x a^x_2
- 2g \bar{A}^\eta_1 \partial_\eta a^y_3 = 0 ,
\\[2mm]
\label{YM-eq-y-E-3-sc2}
&& \Big(  \partial_\tau^2  + \frac{1}{\tau} \,\partial_\tau - \partial_x^2
- \frac{1}{\tau^2} \, \partial_\eta^2 
+ g^2 \tau^2 \bar{A}^\eta_1 \bar{A}^\eta_1 \Big) a^y_3 
+\partial_y \partial_x  a^x_3
+ 2 g \bar{A}^\eta_1 \partial_\eta a^y_2  = 0 ,
\ea
\ba
\label{YM-eq-eta-E-2-sc2}
&& \partial_\eta \big(\partial_x a^x_2 + \partial_y  a^y_2 \big)
+ g \tau^2  \bar{A}^\eta_1 \big( \partial_x a^x_3 + \partial_y  a^y_3\big)  = 0,
\\[2mm]
\label{YM-eq-eta-E-3-sc2}
&& \partial_\eta \big( \partial_x  a^x_3 + \partial_y a^y_3  \big) 
- g \tau^2 \bar{A}^\eta_1 \big(\partial_x  a^x_2 + \partial_y a^y_2 \big) = 0.
\ea

Eqs.~(\ref{YM-eq-tau-E-2-sc2}), (\ref{YM-eq-tau-E-3-sc2}) and (\ref{YM-eq-eta-E-2-sc2}), (\ref{YM-eq-eta-E-3-sc2}) are solved, respectively, if
\be
\partial_x a^x_2 =- \partial_y a^y_2 , 
~~~~~~~~~~~~~~~~
\partial_x a^x_3 =- \partial_y a^y_3 ,
\ee
which substituted into Eqs.~(\ref{YM-eq-x-E-2-sc2}) - (\ref{YM-eq-y-E-3-sc2}) give
\ba
 \label{YM-eq-x-E-2-sc2-2}
&&  \Big( \partial_\tau^2 + \frac{1}{\tau} \, \partial_\tau 
- \partial_x^2 - \partial_y^2 - \frac{1}{\tau^2} \, \partial_\eta^2 
+ g^2\tau^2 \bar{A}^\eta_1 \bar{A}^\eta_1\Big) a^x_2
- 2 g \bar{A}^\eta_1\partial_\eta a^x_3  = 0,
\\[2mm] 
\label{YM-eq-x-E-3-sc2-2}
&&  \Big( \partial_\tau^2  + \frac{1}{\tau} \, \partial_\tau 
- \partial_x^2 - \partial_y^2 - \frac{1}{\tau^2} \, \partial_\eta^2 
+ g^2\tau^2 \bar{A}^\eta_1 \bar{A}^\eta_1 \Big)  a^x_3
+ 2g \bar{A}^\eta_1 \partial_\eta  a^x_2 = 0,
\ea
\ba
 \label{YM-eq-y-E-2-sc2-2}
&&  \Big( \partial_\tau^2  + \frac{1}{\tau} \, \partial_\tau 
- \partial_x^2 - \partial_y^2 - \frac{1}{\tau^2} \, \partial_\eta^2 
+ g^2 \tau^2 \bar{A}^\eta_1 \bar{A}^\eta_1 \Big) a^y_2 
- 2g \bar{A}^\eta_1 \partial_\eta a^y_3 = 0 ,
\\[2mm]
\label{YM-eq-y-E-3-sc2-2}
&& \Big(  \partial_\tau^2  + \frac{1}{\tau} \,\partial_\tau 
- \partial_x^2 - \partial_y^2 - \frac{1}{\tau^2} \, \partial_\eta^2 
+ g^2 \tau^2 \bar{A}^\eta_1 \bar{A}^\eta_1 \Big) a^y_3 
+ 2 g \bar{A}^\eta_1 \partial_\eta a^y_2  = 0 .
\ea

Introducing the functions
\be
X^\pm \equiv a^x_2 \pm i a^x_3, 
~~~~~~~~~~~~~~
Y^\pm \equiv a^y_2 \pm i a^y_3, 
\ee
Eqs.~(\ref{YM-eq-x-E-2-sc2-2}) - (\ref{YM-eq-y-E-3-sc2-2}) are diagonalized as
\ba
\label{eq-X-E}
&&  \Big( \partial_\tau^2 + \frac{1}{\tau} \, \partial_\tau 
- \partial_x^2 - \partial_y^2 - \frac{1}{\tau^2} \, \partial_\eta^2 
\mp i g E \partial_\eta + \frac{1}{4} g^2\tau^2 E^2 \Big) X^\pm =0 ,
\\[2mm]
\label{eq-Y-E}
&&  \Big( \partial_\tau^2 + \frac{1}{\tau} \, \partial_\tau 
- \partial_x^2 - \partial_y^2 - \frac{1}{\tau^2} \, \partial_\eta^2 
\mp i g E \partial_\eta + \frac{1}{4} g^2\tau^2 E^2 \Big) Y^\pm =0 ,
\ea
where we put $\bar{A}^\eta_1 = -\frac{1}{2} E$. Since the equations of $X^\pm$ and $Y^\pm$ are identical, further on we discuss only Eq.~(\ref{eq-X-E}).

Assuming that the functions $X^\pm$ depend on $x, y$ and $\eta$ through $e^{i(k_x x+ k_y y +\nu \eta )}$, Eq.~(\ref{eq-X-E}) becomes
\be
\label{eq-X-E-k-0}
\Big( \partial_\tau^2 + \frac{1}{\tau} \, \partial_\tau 
+ k_x^2 + k_y^2 + \frac{\nu^2}{\tau^2} 
\pm g E \nu + \frac{1}{4} g^2\tau^2 E^2 \Big) X^\pm =0 ,
\ee
which is rewritten as
\be
\label{eq-X-E-k}
\bigg( \partial_\tau^2 + \frac{1}{\tau} \, \partial_\tau 
+ k_T^2 +  \frac{1}{\tau^2}\Big(\nu \pm \frac{1}{2} g \tau^2 E \Big)^2 \bigg) X^\pm =0 .
\ee
where $k_T^2 = k_x^2 + k_y^2$. In the short time limit when $\tau^2 \ll 2\nu/(gE)$, we deal with the Bessel equation of imaginary order $i\nu$ which is solved by the oscillatory function $J_{i\nu}(k_T \tau)$. In the long $\tau$ limit when $\tau^2 \gg 1/(gE)$, $\tau^2 \gg 2\nu/(gE)$ and $\tau^2 \gg k_T^2/(g^2 E^2)$, the equation (\ref{eq-X-E-k}) becomes independent of $\nu$ and gets the form
\be
\label{eq-X-E-k-long-tau}
\Big( \partial_\tau^2 + \frac{1}{4} g^2 E^2 \tau^2  \Big) X^\pm =0 .
\ee
The solution is $X^\pm \sim \exp\big(\pm \frac{i}{4} g E \, \tau^2\big)$ which oscillates with the period decreasing to zero as $\tau \to \infty$. So, the solutions of Eq.~(\ref{eq-X-E-k}), which actually represent waves running away along to axis $z$ to plus and minus infinity, are stable. 

\section{Temporal evolution of glasma background field}
\label{sec-glasma-evolution}

The glasma fields are not stationary but they evolve in time. Consequently, our stability analysis is reliable if a rate of change of the background field is significantly smaller than the growth rate of instability found in Sec.~\ref{sec-stab-B}. To check the condition we consider the evolution of glasma fields using the proper time expansion \cite{Fries:2006pv,Chen:2015wia}. The potentials $\alpha(\tau,{\bf x}_\perp)$ and $\boldsymbol{\alpha}_\perp(\tau, {\bf x}_\perp)$ are expanded in the proper time $\tau$ as
\ba
\label{expansion-1}
\alpha(\tau,{\bf x}_\perp) &=& \alpha^{(0)}({\bf x}_\perp) 
+ \tau \alpha^{(1)}({\bf x}_\perp) 
+ \tau^2 \alpha^{(2)}({\bf x}_\perp) + \cdots ,
\\[2mm]
\label{expansion-2}
\boldsymbol{\alpha}_\perp(\tau,{\bf x}_\perp)&=& \boldsymbol{\alpha}^{(0)}_\perp({\bf x}_\perp) 
+ \tau  \boldsymbol{\alpha}^{(1)}_\perp({\bf x}_\perp) 
+ \tau^2  \boldsymbol{\alpha}^{(2)}_\perp({\bf x}_\perp) + \cdots.
\ea
The zeroth order functions are given by the boundary conditions (\ref{cond1}) and (\ref{cond2}) that is $\alpha_\perp^{(0)}({\bf x}_\perp) = \alpha_\perp(0,{\bf x}_\perp)$ and $\alpha_\perp^{i(0)}({\bf x}_\perp) = \alpha_\perp^i(0,{\bf x}_\perp)$ with $i,j=x,y$. 

One shows that the coefficients multiplying odd powers of $\tau$ in the series (\ref{expansion-1}) and (\ref{expansion-2}) vanish \cite{Chen:2015wia} while the second order coefficients in the fundamental representation  are given as
\ba
\label{al-2}
&& \alpha^{(2)} = \frac{1}{8} \big[{\cal D}^j,[{\cal D}^j,\alpha^{(0)}]\big] ,
\\[2mm]
\label{al-p2}
&& \alpha^{i(2)}_{\perp} = \frac{1}{4} \epsilon^{zij} [{\cal D}^j, B] ,
\ea
where all quantities are taken at the same point ${\bf x}_\perp$ and ${\cal D}^i \equiv \partial^i-ig \alpha_\perp^{i(0)}$. The coefficients (\ref{al-2}) and (\ref{al-p2}) are expressed through the pre-collision potentials $\beta^i_1,~\beta^i_2$ in the following way
\ba
\nn
\alpha^{(2)} &=& \frac{g}{16} \Big(
 - i \partial^j  \partial^j [\beta^i_1,\beta^i_2] 
- g  \partial^j  \big[\beta^j_1 , [\beta^i_1,\beta^i_2] \big]
-  g \partial^j  \big[\beta^j_2, [\beta^i_1,\beta^i_2] \big] \Big] 
\\ [2mm] \label{al-2-final}
&&- g  \big[\beta^j_1 + \beta^j_2, \partial^j [\beta^i_1,\beta^i_2] \big]
+  ig^2 \big[\beta^j_1 + \beta^j_2, \big[\beta^j_1[\beta^i_1,\beta^i_2] \big]
+  ig^2 \big[\beta^j_1 + \beta^j_2, \big[\beta^j_2[\beta^i_1,\beta^i_2] \big]  \Big) ,
\\[2mm]
\label{al-p2-final}
\alpha^{i(2)}_{\perp} &=& \frac{g}{4} \epsilon^{zij}\epsilon^{zkl} \Big( i \partial^j [\beta^k_1,\beta^l_2] 
+g \big[\beta^j_1 + \beta^j_2 , [\beta^k_1,\beta^l_2] \big]\Big) .
\ea

When $\beta^i_1$ and $\beta^i_2$ are independent of ${\bf x}_\perp$, the second order contributions to $\alpha$ and $\boldsymbol{\alpha}_\perp$ are
\ba
\alpha^{(2)} &=& \frac{ig^3}{16} \Big( \big[\beta^j_1 + \beta^j_2, \big[\beta^j_1[\beta^i_1,\beta^i_2] \big]
+  \big[\beta^j_1 + \beta^j_2, \big[\beta^j_2[\beta^i_1,\beta^i_2] \big] \Big) ,
\\[2mm]
 \alpha^{i(2)}_{\perp} &=& \frac{g^2}{4} \epsilon^{zij} \epsilon^{zkl}  
 \big[\beta^j_1 + \beta^j_2 , [\beta^k_1,\beta^l_2] \big] .
\ea

Now, let us assume that the pre-collision potentials, which are purely transverse, are $\boldsymbol{\beta}_1 = (\beta_1,0)$ and $\boldsymbol{\beta}_2 = (0,\beta_2)$. Then, $[\beta^i_1,\beta^i_2]$ vanishes and so does the initial electric field (\ref{Ez-initial}). The initial magnetic field (\ref{Bz-initial}) equals $B = ig [\beta_1,\beta_2]$. The second order contributions are 
\ba
\alpha^{(2)} &=& 0,
\\[2mm]
 \alpha^{x(2)}_{\perp} &=& \frac{g^2}{4} \epsilon^{zxy} \epsilon^{zxy}  
 \big[\beta_2 , [\beta_1,\beta_2] \big]
 = \frac{g^2}{4} \big[\beta_2 , [\beta_1,\beta_2] \big] ,
 \\[2mm]
 \alpha^{y(2)}_{\perp} &=& \frac{g^2}{4} \epsilon^{zyx} \epsilon^{zxy}  
 \big[\beta_1, [\beta_1,\beta_2] \big] 
= -\frac{g^2}{4} \big[\beta_1, [\beta_1,\beta_2] \big] .
\ea

Going to the adjoint representation of the SU(2) group, on finds
\ba
\alpha^{x(2)}_{\perp a} &=&\frac{g^2}{4}(\beta_2^b \beta_1^b \beta_2^a - \beta_2^b \beta_1^a \beta_2^b ),
 \\[2mm]
\alpha^{y(2)}_{\perp a} &=& -\frac{g^2}{4}  (\beta_1^b \beta_1^b \beta_2^a - \beta_1^b \beta_1^a \beta_2^b ).
\ea

If $\beta_1^a = \lambda^{-1} \delta^{a3} \sqrt{B/g}$ and $\beta_2^a = \lambda \delta^{a2} \sqrt{B/g}$, the second order contributions to $\boldsymbol{\alpha}_{\perp a}$ are
\ba
\alpha^{x(2)}_{\perp a} &=&- \frac{1}{4} \delta^{a3} \lambda g^{1/2} B^{3/2},
 \\[2mm]
\alpha^{y(2)}_{\perp a} &=& -\frac{1}{4} \delta^{a2} \lambda^{-1} g^{1/2} B^{3/2}.
\ea

Taking into account the 0th and 2nd order contributions in proper time expansion, the function $\alpha_a$ and the $x$ and $y$ components of the function $\boldsymbol{\alpha}_{\perp a}$ are equal to
\ba
\nn
\alpha_a &=&  {\cal O}(\tau^4) ,
\\[2mm]
\label{potential-2nd-order}
\alpha_{\perp a}^x &=&   \delta^{a3}  \lambda^{-1} \sqrt{B/g} 
\Big( 1 - \frac{1}{4} \lambda^2 g B \tau^2 + {\cal O}(\tau^4)\Big) ,
\\[2mm] \nn
\alpha_{\perp a}^y &=&   \delta^{a2} \lambda \sqrt{B/g} 
\Big(1 - \frac{1}{4} \lambda^{-2} g B \tau^2 + {\cal O}(\tau^4)\Big) . 
\ea
One observes that the potential (\ref{potential-2nd-order}) generates the zeroth and second order longitudinal magnetic field and the first order transverse electric field.

\section{Discussion and conclusions}
\label{sec-discussion}

In our stability analysis the electric and magnetic fields are assumed to be space-time uniform. However, the glasma fields generated at the earliest phase of ultrarelativistic heavy-ion collisions are not uniform, neither spatially nor temporally. So, one asks to what extent our results apply to the description of real glasma.

The correlators of glasma fields, discussed {\it e.g.} in Sec. IIID of \cite{Carrington:2022bnv}, show that the fields are spatially uniform in the transverse plane at a scale $L$ which is in between $Q_s^{-1}$ and $\Lambda_{\rm QCD}^{-1}$ where $Q_s \approx 2~{\rm GeV}$ is the saturation scale and $\Lambda_{\rm QCD} \approx 0.2~{\rm GeV}$ is the QCD confinement scale at which color charges are neutralized. Assuming that the domain, where the field is uniform, is a square centered at ${\bf r}_\perp=0$ and demanding that the real potentials $a_a^x$ and $a_a^y$ vanish at the edge of the square, the wave vectors $k_x, k_y$ should be replaced as
\be
\label{replacement}
(k_x,k_y) \longrightarrow (2l_x +1, 2l_y +1) \frac{\pi}{L} , 
\ee
where $l_x, l_y$ are integer numbers. Consequently, a spectrum of eigenmodes becomes discrete and the unstable mode found in Sec.~\ref{sec-stab-B} can disappear if the minimal momentum $\pi/L$ is sufficient to stabilize it. Using the growth rate of the instability estimated as $\sqrt{gB - \frac{1}{3}k_T^2}$, one finds that due to the replacement (\ref{replacement}) the instability appears if 
\be
\label{insta-condition}
gB  >  \frac{2\pi^2}{3L^2}.
\ee 
Since $g\approx 1$ and $B \approx Q_s^2 \approx 4~{\rm GeV}^2$, the condition (\ref{insta-condition}) is satisfied for $L^{-1} \approx \Lambda_{\rm QCD} \approx 0.2~{\rm GeV}$. Taking into account that the generation of chromodynamic fields in heavy-ion collisions is a random process and consequently, a magnitude of the field and a size of the domain, where the field is uniform, vary both in an individual collision and from collision to collision, we expect that the condition (\ref{insta-condition}) is not always satisfied but it often is and then, the unstable mode occurs. 

The initial glasma fields are not stationary and as discussed in Sec.~\ref{sec-glasma-evolution} the magnetic field changes with the characteristic rate $g B \tau$ where we put $\lambda =1$. It is smaller than the instability growth rate estimated as $\sqrt{gB}$ when $\tau < (gB)^{-1/2}$  but it is bigger when $\tau > (gB)^{-1/2}$. It suggests that the initial magnetic field can be treated as stationary only for a very short time. However, one should take into account that our estimate of the field rate of change is obtained in the second order of the proper time expansion. The calculations using the proper time expansion, which are presented in {\it e.g.} \cite{Carrington:2022bnv}, show that there are usually alternating signs of successive terms in the proper time expansion and consequently, the temporal evolution is significantly slower than a second order result suggests. Therefore, we expect that the initial magnetic field can be treated as stationary for time scale $(gB)^{-1/2}$ or even longer. 

Let us now confront our findings with results of the simulations of glasma evolution \cite{Romatschke:2005pm,Romatschke:2006nk} which actually were the main motivation of our work. The simulations showed that the  glasma is unstable and the instability was identified with the Weibel mode \cite{Romatschke:2005pm,Romatschke:2006nk}. We first note that the fastest unstable mode found in \cite{Romatschke:2005pm,Romatschke:2006nk} grows like $e^{\sqrt{\tau}}$ while that of the glasma initial magnetic field as  $e^{\tau}$. The discrepancy can be removed taking into account that the longitudinal magnetic field decreases, as our formulas (\ref{potential-2nd-order}) show. However, we are not going to pursue this path as there are more important reasons not to interpret results of the simulations \cite{Romatschke:2005pm,Romatschke:2006nk} as due to the glasma initial field instability. 

When the fastest growing mode found in \cite{Romatschke:2005pm,Romatschke:2006nk} is fitted with $e^{\gamma \tau}$, the maximal growth rate is $\gamma \approx 0.00272 \, g^2 \mu$ where $\mu$ is the surface density of color charges of incoming nuclei\footnote{We have taken into account the factor of 2 as the growth of the energy-momentum tensor not the growth of the field was obtained in \cite{Romatschke:2005pm,Romatschke:2006nk}.}. Using the authors' estimate $g^2 \mu \approx 20~{\rm fm}^{-1}$ for LHC, we get $\gamma \approx 0.05 ~{\rm fm}^{-1}$.  The maximal growth rate of the unstable mode of the glasma initial field is $\sqrt{gB} \approx  Q_s$ and it occurs for $k_T =0$ and $\lambda =1$. Estimating the saturation scale as $Q_s = 2$ GeV, the maximal growth rate is $\sqrt{gB} \approx 10 ~{\rm fm}^{-1}$ which is 200 times bigger than that from \cite{Romatschke:2005pm,Romatschke:2006nk}. Although, the growth rate is smaller for $k_T > 0$ and/or $\lambda \not=1$, it is hard to expect that the two very different growth rates describe the same physical phenomenon. 

In the glasma simulations \cite{Romatschke:2005pm,Romatschke:2006nk} the instability shows up only if the initial condition includes fluctuations which depend on the space-time rapidity $\eta$ and consequently violate the boost invariance. This is actually the crucial argument to identify the instability as the Weibel mode which requires a finite longitudinal momentum \cite{Mrowczynski:2016etf}. The unstable mode of the initial glasma field can occur at any $\nu$ including $\nu=0$ which corresponds to the boost invariant configuration. While the growth rate is independent of $\nu$, the modes start growing at $\tau=\nu/\sqrt{gB}$. The mode with $\nu=0$, which is independent of $\eta$, starts with no delay and consequently, it is dynamically most important. 

The instabilities of initial glasma fields we have studied analytically here are presumably responsible for a rapid temporal evolution of glasma field correlators investigated in \cite{Ruggieri:2017ioa} using numerical simulations. The authors of  Ref.~\cite{Ruggieri:2017ioa} found that the correlator of chromomagentic fields, which are initially uniform, changes with a characteristic time of the order $Q_s^{-1}$. However, a more detailed analysis is necessary to confirm the supposition. 

We conclude our considerations as follows. The initial glasma field configuration is unstable if the fields are sufficiently uniform both spatially and temporally. Since the process of generation of chromodynamic fields in heavy-ion collisions is stochastic and the field's characteristics fluctuate we expect that the condition of uniformity is often satisfied. The time of the instability development is of order 0.1 fm/$c$ which is much shorter than that of the instability found in the glasma simulations \cite{Romatschke:2005pm,Romatschke:2006nk} which is of order 10 fm/$c$. The fastest unstable mode of the initial glasma field is boost invariant in contrast to the Weibel mode advocated in \cite{Romatschke:2005pm,Romatschke:2006nk} which requires breaking of the boost invariance. So, we conclude that the instability found in the study \cite{Romatschke:2005pm,Romatschke:2006nk} is not the instability of the glasma initial field. To observe the initial glasma field instability, if it indeed occurs, one needs a glasma simulation of high temporal resolution, much higher than that from  \cite{Romatschke:2005pm,Romatschke:2006nk}.

\section*{Acknowledgments}

This work was partially supported by the National Science Centre, Poland under grant 2018/29/B/ST2/00646.

\appendix*
\section{Bessel equations}

The Bessel equation is
\be
\label{Bessel-eq}
\Big(\frac{d^2}{dx^2} + \frac{1}{x}\frac{d}{dx} + 1 - \frac{\alpha^2}{x^2}\Big) f(x) = 0 ,
\ee
and its two linearly independent solutions are the Bessel functions 
\ba
J_\alpha(x) &\equiv& \sum_{m =0}^\infty \frac{(-1)^m}{m! \,\Gamma(m + \alpha + 1)} 
\Big(\frac{x}{2}\Big)^{2m + \alpha},
\\[2mm]
Y_\alpha(x) &\equiv& \frac{J_\alpha(x) \, \cos(\alpha \pi) - J_{-\alpha}(x)}{\sin(\alpha \pi)} ,
\ea
which oscillate around zero for $x \in \mathds{R}$. At $x=0$ the functions $J_\alpha(x)$ are finite but $Y_\alpha(x)$ diverge. For $x \gg |\alpha^2 - \frac{1}{4}|$ the following approximation holds
\be
J_\alpha(x) = \sqrt{\frac{2}{\pi x}} \, \cos\Big( x - \frac{\pi}{2}\alpha - \frac{\pi}{4}\Big) + {\cal O}(x^{-3/2}).
\ee

Changing the variable $x= it$, Eq.~(\ref{Bessel-eq}) becomes the modified Bessel equation
\be
\label{Bessel-eq-mod}
\Big(\frac{d^2}{dt^2} + \frac{1}{t}\frac{d}{dt} - 1 - \frac{\alpha^2}{t^2}\Big) g(t) = 0 ,
\ee
where $g(t) = f(it)$ and the two linearly independent solutions are the modified Bessel functions 
\ba
I_\alpha(t) &\equiv& i^{-\alpha} J_\alpha(it) =\sum_{m =0}^\infty \frac{1}{m! \,\Gamma(m + \alpha + 1)} 
\Big(\frac{t}{2}\Big)^{2m + \alpha},
\\[2mm]
K_\alpha(t) &\equiv& \frac{\pi}{2} \frac{I_{-\alpha}(t) - I_{\alpha}(t)}{\sin(\alpha \pi)} .
\ea
The functions $I_\alpha(t)$ are finite at $t=0$ and exponentially grow with $t$ for $t \in \mathds{R}$. The functions $K_\alpha(t)$ are infinite at $t=0$ and exponentially decay as $t$ grows.  For $t \gg |\alpha^2 - \frac{1}{4}|$ we have the approximation 
\be
I_\alpha(t) = \frac{e^t}{\sqrt{\pi t}} \big( 1 + {\cal O}(t^{-1}) \big).
\ee

Changing the variable $x = a\tau$ in Eq.~(\ref{Bessel-eq}) and  the variable $t = a\tau$ in Eq.~(\ref{Bessel-eq-mod}), the Bessel and modified Bessel equations read
\ba
\label{Bessel-eq-2}
\Big(\frac{d^2}{d\tau^2} + \frac{1}{\tau}\frac{d}{d\tau} + a^2 - \frac{\alpha^2}{\tau^2}\Big) h(\tau) = 0,
\\[2mm]
\label{Bessel-eq-mod-2}
\Big(\frac{d^2}{d\tau^2} + \frac{1}{\tau}\frac{d}{d\tau} - a^2 - \frac{\alpha^2}{\tau^2}\Big) h(\tau) = 0 ,
\ea
where $h(\tau) \equiv f(a\tau)$ or $h(\tau) \equiv g(a\tau)$. The solutions are $J_\alpha(a \tau), Y_\alpha(a \tau)$ and $I_\alpha(a \tau), K_\alpha(a \tau)$, respectively.


\end{document}